\documentclass{aa}  
\usepackage{graphicx}
\usepackage{amsmath}	
\usepackage{amssymb}	
\usepackage{relsize}
\usepackage{txfonts}

\newcommand{\kms}{\hbox{km s$^{-1}$}}

\begin{document}

   \title{The Fornax 3D project: a two-dimensional view of the stellar initial mass function in the massive lenticular galaxy FCC\,167}

   \subtitle{}

   \author{
I. Mart\'in-Navarro \inst{1,2}, M. Lyubenova \inst{3}, G. van de Ven \inst{3,4}, J. Falc\'on-Barroso \inst{5,6}, L.~Coccato\inst{3}, E.~M.~Corsini\inst{7, 8}, D.~A.~Gadotti\inst{3}, E.~Iodice\inst{9},  F. La Barbera\inst{9}, R.~M.~McDermid\inst{10, 11}, F.~Pinna\inst{5,6}, M.~Sarzi\inst{12,13}, S. Viaene\inst{13,14}, P.~T.~de~Zeeuw\inst{15,16}, \and L. Zhu\inst{2}
          }

   \institute{
University of California Observatories, 1156 High Street, Santa Cruz, CA 95064, USA\\
\email{imartinn@ucsc.edu}
\and
Max-Planck Institut f\"ur Astronomie, Konigstuhl 17, D-69117 Heidelberg, Germany
\and
European Southern Observatory, Karl-Schwarzschild-Str. 2, 85748 Garching b. M\"unchen, Germany
\and
University of Vienna, Department of Astrophysics, T\"urkenschanzstrasse 17, 1180 Vienna, Austria
\and
Instituto de Astrof\'{\i}sica de Canarias,c/ V\'{\i}a L\'actea s/n, E38205 - La Laguna, Tenerife, Spain
\and
Departamento de Astrof\'isica, Universidad de La Laguna, E-38205 La Laguna, Tenerife, Spain
\and
Dipartimento di Fisica e Astronomia `G. Galilei', Universit\`a di Padova, vicolo dell'Osservatorio 3, I-35122 Padova, Italy
\and 
INAF--Osservatorio Astronomico di Padova, vicolo dell'Osservatorio 5, I-35122 Padova, Italy
\and
INAF--Osservatorio Astronomico di Capodimonte, via Moiariello 16, I-80131 Napoli, Italy
\and
Department of Physics and Astronomy, Macquarie University, Sydney, NSW 2109, Australia
\and 
Australian Astronomical Observatory, PO Box 915, Sydney, NSW 1670, Australia
\and
Armagh Observatory and Planetarium, College Hill, Armagh, BT61 9DG, UK
\and
Centre for Astrophysics Research, University of Hertfordshire, College Lane, Hatfield AL10 9AB, UK
\and
Sterrenkundig Observatorium, Universiteit Gent, Krijgslaan 281, B-9000 Gent, Belgium
\and 
Sterrewacht Leiden, Leiden University, Postbus 9513, 2300 RA Leiden, The Netherlands
\and 
Max-Planck-Institut fuer extraterrestrische Physik, Giessenbachstrasse, 85741 Garching bei Muenchen, Germany
             }

   \date{Received; accepted}
   
   \titlerunning{A two-dimensional view of the IMF}
\authorrunning{Mart\'in-Navarro et al.}  
 
  \abstract
   {
   The stellar initial mass function (IMF) regulates the baryonic cycle within galaxies, and is a key ingredient to translate observations into physical quantities. Although for decades it was assumed to be universal, there is now growing observational evidence showing that the center of massive early-type galaxies host an enhanced population of low-mass stars compared to the expectations from the Milky Way. Moreover, these variations in the IMF have been found to be related to the radial metallicity variations in massive galaxies. We present here a two-dimensional stellar population analysis of the massive lenticular galaxy FCC\,167 (NGC\,1380) as part of the Fornax3D project. Using a newly developed stellar population fitting scheme, we derive a full two-dimensional IMF map of an early-type galaxy. This two-dimensional analysis allows us go further than a radial analysis, showing how the metallicity changes along a disc-like structure while the IMF follows a distinct, less disky distribution. Thus, our findings indicate that metallicity cannot be the sole driver of the observed radial IMF variations. In addition, a comparison with the orbital decomposition shows suggestive evidence of a coupling between stellar population properties and the internal dynamical structure of FCC\,167, where metallicity and IMF maps seem to track the distribution of cold and warm orbits, respectively.
   }

\keywords{galaxies: formation -- galaxies: evolution -- galaxies: fundamental parameters -- galaxies: stellar content -- galaxies: elliptical}

   \maketitle
%

\section{Introduction}

The stellar initial mass function (IMF) describes the mass spectrum of stars at birth and plays a fundamental role in our understanding of galaxies. From the stellar feedback \citep[e.g.][]{Thales18,Barber18} to the chemical enrichment \citep[e.g.][]{Ferreras15,MN16,Philcox18,Barber18b}, baryonic processes in the Universe ultimately depend on the IMF. Moreover, observationally, our interpretation of the electromagnetic spectrum of unresolved stellar populations is heavily sensitive to the IMF. Beyond a few Mpc, where the properties of resolved individual stars cannot still be measured, translating spectro-photometric measurements into physical quantities requires strong assumptions on the shape of the IMF. From star formation rate \citep[e.g.][]{Kennicutt98,Madau14} and stellar mass measurements \citep[e.g.][]{Mitchell13,Courteau14,Bernardi18} to chemical enrichment predictions \citep[e.g.][]{McGee14,Clauwens16}, the shape of the IMF has to be either fixed or modelled. 

The pioneering work of \citet{Salp:55} showed that the IMF in the Milky Way can be described by a power law
\begin{equation} \label{eq:imf}
\Phi(\log M) = d N / d \log M \propto M^{-\Gamma}
\end{equation}

\noindent with a slope $\Gamma = 1.35$ for stellar masses above 1M$_\odot$. These initial measurements were later extended to lower-mass stars (M$\lesssim 0.5$M$_\odot$), where the slope of the IMF was found to be flatter  ($\Gamma\sim0$) than for massive stars \citep[e.g.][]{Miller79}. The seminal works of \citet{mw,Kroupa} and \citet{Chabrier} consolidated the idea of a universal, Milky Way-like IMF, independent of the local star formation conditions \citep[e.g.][]{bastian}.

IMF measurements are, however, not limited to the relatively small local volume around the Milky Way. In unresolved stellar populations, two main approaches have been developed to constrain the IMF shape. The first approach makes use of the effect of the IMF on the mass-to-light (M/L) ratio. Low-mass stars constitute the bulk of stellar mass in galaxies, but high-mass stars dominate the light budget. Hence, a change in the IMF slope (i.e. in the relative number of low-mass to massive stars) will have a measurable impact on the expected M/L, and the shape of the IMF can be estimated by independently measuring the total mass (through dynamics or gravitational lensing) and luminosity of a stellar population \citep[see e.g][]{Treu,thomas11}. This technique, however, suffers from a strong degeneracy between dark matter halo mass and IMF slope \citep[e.g.][]{auger}. Alternatively, the shape of the IMF can also be measured in unresolved stellar populations by analyzing their integrated absorption spectra. In particular, IMF-sensitive features subtly vary, at fixed effective temperature, with the surface gravity of stars, and therefore can be used to measure the dwarf-to-giants ratio, i.e., the slope of the IMF \citep[e.g][]{vandokkum}. Although more direct than the dynamical/lensing approach, measuring the IMF from integrated spectra is observationally challenging as the low-mass star contribution to the observed spectrum is usually outshined by the flux emitted by more massive stars. 

The relative simplicity of the stellar populations (i.e. roughly coeval) in massive early-type galaxies (ETGs) has made them benchmark test cases to study IMF variations beyond the Local Group. In addition, massive ETGs are more metal-rich, denser and have experienced more intense star formation events than the Milky Way, and thus, the universality of the IMF shape can be tested under much more extreme conditions. Over the last decade, IMF studies in ETGs have consistently indicated a non-universal IMF shape in massive ETGs, as the IMF slope becomes steeper (i.e. with a larger fraction of low-mass stars) with increasing galaxy mass. It is important to note that ETGs host very old stellar populations and thus, IMF measurements in these objects are restricted to long-lived stars, with masses $m\lesssim1$M$_\odot$, and therefore insensitive to variations in the high-mass end of the IMF.
In general, the agreement between dynamics-based \citep{auger,cappellari,Dutton12,wegner12,Tortora13b,Lasker13,Tortora14,Corsini17} and stellar population-based \citep{vandokkum,spiniello12,Spiniello2013,Spiniello15,conroy12,Smith12,ferreras,labarbera,Tang17} studies support a systematic variation in the IMF of massive ETGs. However, dynamical and stellar population studies do not necessarily agree on the details \citep[see e.g.][]{smith,Smith13,Newman17}. This seems to suggest that both approaches might be effectively probing different mass-scales of the IMF, although differences between dynamical and stellar population-bases studies are not as striking after a proper systematics modeling \citep{Lyubenova16}.

IMF variations with galaxy mass, however, provide limited information about the process(es) shaping the IMF as, in general, galaxy properties tend to scale with galaxy mass. Therefore, galaxy-wide IMF variations may be equally attributed to a number of different mechanisms \citep[e.g.][]{conroy12,labarbera,LB15}. This observational degeneracy can be partially broken by analyzing how the IMF shape changes as a function of radius, since different parameters like velocity dispersion, stellar density, metallicity or abundance pattern vary differently with galactocentric distance. Since first measured in the massive galaxy NGC\,4552 \citep{MN15a}, radial IMF gradients have been widely found in a large number of massive ETGs \citep{MN15b,LB16,Davis17,vdk17,Oldham,Parikh,Sarzi18,Vaughan18a}. These spatially resolved IMF studies have shown how IMF variations occur in the inner regions of massive ETGs, and that metallicity seems to be the local property that better tracks the observed IMF variations \citep{MN15c}. However, it is not clear whether the observed correlation with metallicity is sufficient to explain all IMF variations \citep[e.g.][]{Alexa17}, or even if there are massive ETGs with Milky Way-like IMF slopes \citep{McConnell16,Zieleniewski17,Alton18,Vaughan18b}.

The aim of this work is to take a step further in the observational characterization of the IMF by analyzing its two-dimensional (2D) variation in the massive ($M_B=-20.3$), fast-rotating ETG FCC\,167 (NGC\,1380), as part of the Fornax 3D project (F3D). Taking advantage of the unparalleled capabilities of the  Multi  Unit  Spectroscopic Explorer (MUSE) integral-field spectrograph \citep{Bacon10}, we present here a 2D analysis of the stellar population properties of FCC\,167, showing for the first time the IMF map of a massive ETG. The paper is organized as follows: in \S~\ref{sec:data} we briefly present the data. Stellar population model ingredients are described in \S~\ref{sec:model}. \S~\ref{sec:fif} contains a detailed explanation about the stellar population modeling, and fitting method 9 and in \S~\ref{sec:results} the results of the stellar population analysis of FCC\,167 are presented. In \S~\ref{sec:discussion} we discuss our findings. Finally, in \S~\ref{sec:summary} we summarize the main conclusions of this work, briefly describing future IMF efforts within the Fornax 3D project.

\section{Data} \label{sec:data}

We based our stellar population analysis on MUSE data from F3D described in \citet{f3d}. In short, F3D is an IFU survey of 33 bright ($m_B < 15$) galaxies selected from the Fornax Cluster Catalog \citep[FCC, ][]{Ferguson89} within the virial radius of the Fornax cluster \citep{Drinkwater01}. The survey was carried out using the Wide Field Mode of the MUSE IFU \citep{Bacon10}, which provides a 1$\times$1 arcmin$^2$ field-of-view at a 0.2 arcsec pixel$^{-1}$ spatial scale. The wavelength range covers from 4650\AA \ to 9300\AA, with an spectral sampling of 1.25 \AA \ pixel$^{-1}$ and a nominal resolution of FWHM = 2.5\AA \ at $\lambda=7000$\AA.

This work is focused on the ETG (S0/a) galaxy FCC\,167, \mbox{located} at a distance of 21.2 Mpc \citep{Blakeslee09} and at 222 kpc from  NGC\,1399, the brightest galaxy in the cluster. The total stellar mass of FCC\,167 is $9.85\times10^{10} M_\odot$, with an effective radius of $R_e=6.17$ kpc (60 arcsec) in the $i$ band \citep{Enrica18}. \citet{f3d} found that FCC\,167 has two embedded (thin and thick) discs based on its orbital decomposition. The thin disc is clearly seen in the photometric structure, along with some other interesting features. Nebular gas emission is present in the central regions \citep{Viaene19}, and the exquisite spatial resolution of the MUSE data allows us to systematically detect and characterise planetary nebulae within the MUSE field-of-view \citep{f3d}.

The final F3D data cube of FCC\,167 combines three different pointings, covering from the center of the galaxy out to $\sim$4 R$_e$, with a total exposure time per pointing of $\sim 1.2$ hours. Data reduction was done using the MUSE pipeline \citep{Weilbacher16} running under the ESO Reflex environment \citep{Freudling13}. The initial sky subtraction was done using either dedicated sky exposures or IFU spaxels free from galactic flux at the edge of the MUSE field of view. In a later step, the sky subtraction process was further improved by using the Zurich Atmospheric Purge algorithm \citep{Soto16}. More details on the observational strategy and data reduction process are given in the F3D presentation paper \citep{f3d}.

In order to measure IMF variation in the absorption spectra of galaxies it is necessary to accurately model variations of a few percent in the depth of gravity-sensitive features. We achieved the required precision level by spatially binning the MUSE data into Voronoi bins with a minimum signal-to-noise ratio (S/N) of $\sim$100 per bin \citep{voronoi}. This S/N threshold is similar to that used in previous IMF studies, but the exquisite spatial resolution and sensitivity of MUSE, combined with the 2D information, provides an unprecedented number of spatial bins: while IMF gradients have usually been measured using $\sim$ 10 data points \citep[e.g.][]{MN15a,vdk17}, our S/N=100 binned cube of FCC\,167 consists of more than 6000 independent Voronoi bins.

\section{Stellar population model ingredients} \label{sec:model}

Our stellar population analysis relies on the most recent version of the MILES evolutionary stellar population synthesis models \citep{Vazdekis15}. These models are fed with the MILES stellar library of \citet{Pat06}, with a constant spectral resolution of 2.51 \AA \ \citep[FWHM, ][]{Jesus11}.

The main difference of the \citet{Vazdekis15} single stellar population models (SSP) with respect to previous MILES models \citep[e.g.][]{miles} is the treatment of the $\alpha$-elements abundance ratio. In this new set of models, MILES stars are used to populate BaSTI isochrones \citep{basti1,basti2} which were explicitly calculated at [$\alpha$/Fe]=0.4 and at the solar scale ([$\alpha$/Fe] = 0.0). To compute [$\alpha$/Fe] = 0.0 models at low metallicities, a regime where MILES stars are $\alpha$-enhanced \citep{Milone11}, a theoretical differential correction was applied on top of the fully-empirical SSP. The same procedure was followed to generate [$\alpha$/Fe]=0.4 SSP models at high metallicities. Therefore this new version of the MILES models allows for a self-consistent treatment of the abundance pattern.

In addition to variable [$\alpha$/Fe] (0.0 and +0.4), the \citet{Vazdekis15} MILES models fed with BaSTI cover a range in metallicities ([M/H]) from -2.27 to +0.26\footnote{Note that, although there are MILES/BaSTI models at [M/H]=+0.4, these predictions are not considered {\it safe} and therefore we do not include them as templates in our analysis}, expanding from 0.03 Gyr to 14 Gyr in age. The wavelength range in the [$\alpha$/Fe]-variable models is relatively short \citep{Vazdekis16}, from $\lambda$=3540 \AA \ to $\lambda$=7410 \AA, but it contains enough gravity-sensitive features to safely measure the effect of the IMF.

For the IMF functional form, we assumed the so-called {\it bimodal} shape \citep{vazdekis96}. In this parametrization, 
the IMF is varied through the (logarithmic) high mass end slope $\Gamma_\mathrm{B}$. The main difference between this IMF shape and a single power-law parametrization is that, for masses below $\sim0.5$M$_\odot$, the bimodal IMF flattens. This feature allows for a better agreement with dynamical IMF measurements \citep{Lyubenova16} while recovering a Milky Way-like behavior for $\Gamma_\mathrm{B} = 1.3$. Although $\Gamma_\mathrm{B}$ controls the high mass end slope, the number of low-mass stars is effectively varied at the same time as the integral of the IMF is normalized to 1 M$_\odot$. Note that a variation in the high-mass end of the IMF as presented here would be in tension with the observed chemical composition of massive ETGs, unless the slope of the IMF also changes with time \citep[][]{weidner:13,Ferreras15,MN16}. However, our stellar population analysis of FCC\,167 is completely insensitive to this potential issue (see details below).

\subsection{IMF in quiescent galaxies: the $\xi$ parameter}

Understanding IMF variations in unresolved, old stellar populations from optical and near-IR spectra requires the acknowledgement of two empirical limitations. First, only stars less massive than $\sim1$M$_\odot$ contribute to the light budget and hence, IMF measurements from integrated spectra are insensitive to variations of the high mass end slope. Second, the contribution from very low mass stars close to the Hydrogen burning limit ($m\sim0.1$M$_\odot$) is virtually unconstrained unless very specific near-IR spectral features are targeted \citep{conroy12,conroy17}. The lack of constraints on the number of very low-mass stars explains why a single-power law IMF parametrization fits equally well the observed spectra of massive ETGs \citep{labarbera,Spiniello2013}, but dramatically overestimates the expected M/L ratios \citep{ferreras,Lyubenova16}. In practice, these two limitations imply that stellar population-based IMF measurements are mostly sensitive to the IMF slope for stars with masses 0.2$\lesssim m \lesssim$1M$_\odot$ regardless of the adopted IMF parametrization.  

Given the rising number of stellar population-based IMF measurements and the need for an unbiased comparison among them, we introduce here a new quantity, $\xi$, which is virtually independent of the IMF parametrization. The $\xi$ parameter, quantifying the mass fraction locked in low-mass stars, is defined as follows

\begin{equation}\label{eq:shape}
\xi \equiv \frac{\int_{m=0.2}^{m=0.5} \Phi(\log m) \ dm}{\int_{m=0.2}^{m=1} \Phi(\log m) \ dm} = 
\frac{\int_{m=0.2}^{m=0.5} m \cdot X(m) \ dm}{\int_{m=0.2}^{m=1} m \cdot X(m) \ dm} 
\end{equation}

\noindent where in the second equality the IMF, $X(m)$, is expressed in linear mass units. This is similar to the definition of F$_{0.5}$ in \citet{labarbera} but $\xi$ is normalized only to the mass contained in stars below 1M$_\odot$, while \citet{labarbera} normalized to the mass in stars below 100 M$_\odot$. Hence, $\xi$ does not depend on the number of massive stars, which in fact cannot be measured from the absorption spectra of ETGs. Moreover, since the denominator in Eq.~\ref{eq:shape} is roughly equivalent to the total stellar mass for old stellar populations, $\xi$ offers a quick conversion factor to transform the observed stellar mass of a galaxy into the total mass in low-mass stars. This definition of $\xi$ explicitly takes into account the fact that very low-mass stars (M$\lesssim0.2$M$_\odot$) are not strongly constrained by most optical/near-IR spectroscopic data. Note also that $\xi$ does not account for the amount of mass locked in stellar remnants. This {\it dark} stellar mass is only measurable through dynamical studies, and it heavily depends on the shape and slope of the high-mass end of the IMF.

The $\xi$ parameter, as defined by Eq.~\ref{eq:shape} is therefore a useful quantity to compare different IMF measurements, as shown in \S~\ref{sec:fif}, and it can be even applied to IMF functional forms without a well-defined low-mass end IMF slope \citep[e.g.][]{Chabrier14,conroy17}.  Table~\ref{table:imf} shows how commonly used IMF functional forms can be translated into $\xi$ mass ratios. 

\begin{table}
\caption{\label{table:imf} Conversion coefficients between commonly used IMF shapes and $\xi$ mass ratios. For a given IMF slope $\gamma$, the corresponding $\xi$ can be accurately approximated by a polynomial $\xi(\gamma) = c_0 + c_1\gamma +  c_2\gamma^2 + c_3\gamma^3$}
\centering
\begin{tabular}{cccc}
\multicolumn{4}{c}{Single-power law ($\Gamma$)\tablefootmark{a}} \\
\hline\hline
$c_0$ & $c_1$ & $c_2$ & $c_3$ \\
\hline
0.3751 & 0.1813 & 0.0223 & -0.0095 \\
\multicolumn{4}{c}{ } \\
\multicolumn{4}{c}{Broken-power law ($\Gamma_\mathrm{B}$)} \\
\hline\hline
$c_0$ & $c_1$ & $c_2$ & $c_3$ \\
\hline
0.3739 & 0.1269 & 0.0000 & -0.0014 \\
\hline
\end{tabular}
\tablefoot{For a \citet{Chabrier} IMF, $\xi=0.4607$; for \citet{Kroupa}, $\xi=0.5194$, and for \citet{Salp:55}, $\xi=0.6370$ \\
\tablefoottext{a}{$\Gamma$ is in $\log$ units. In linear mass units, the IMF slope is $\alpha=\Gamma+1$.}
}
\end{table}

\section{Full-Index-Fitting: a novel approach}~\label{sec:fif}

\begin{figure*}
\centering
\includegraphics[width=9cm]{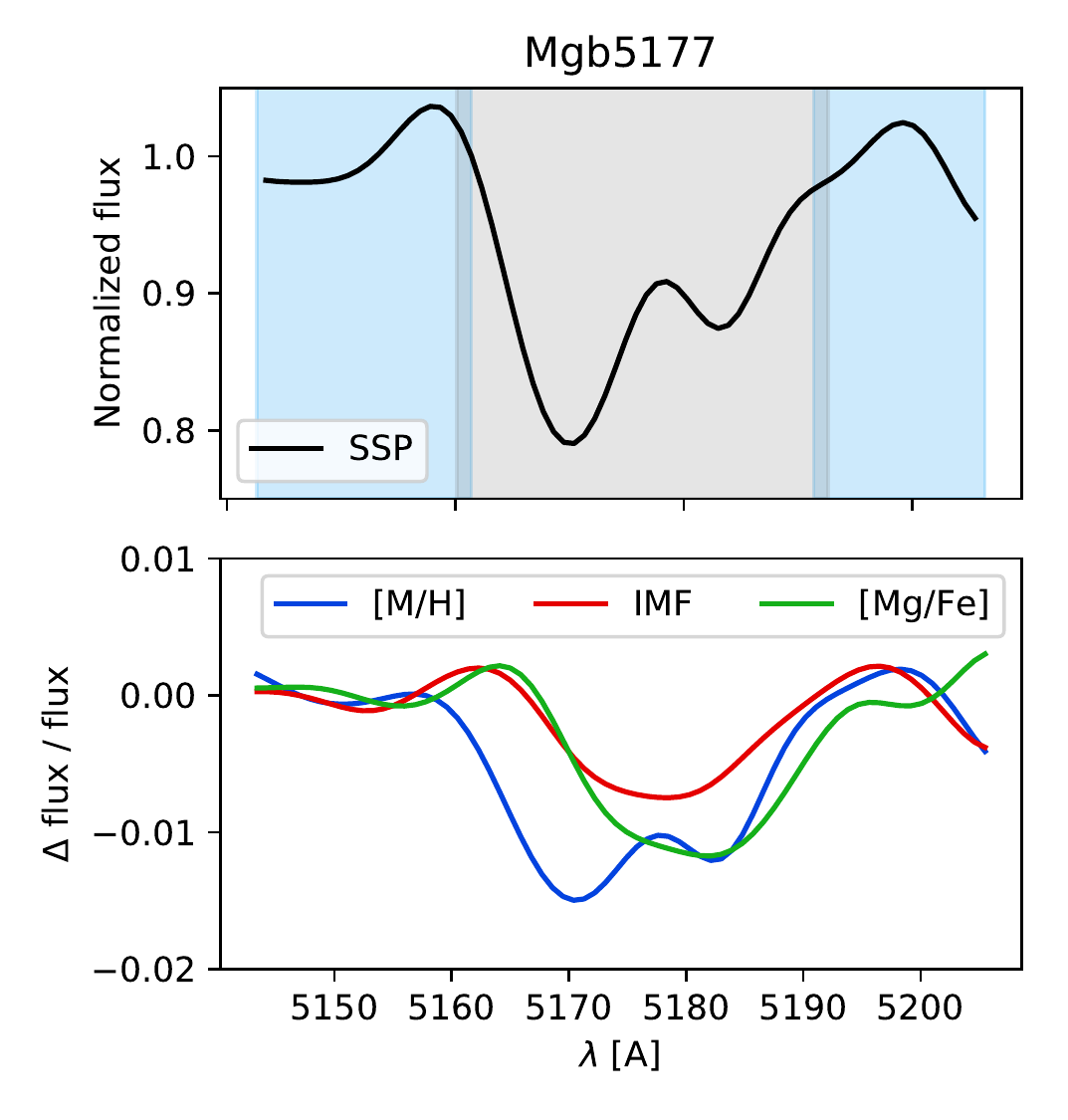}
\includegraphics[width=9cm]{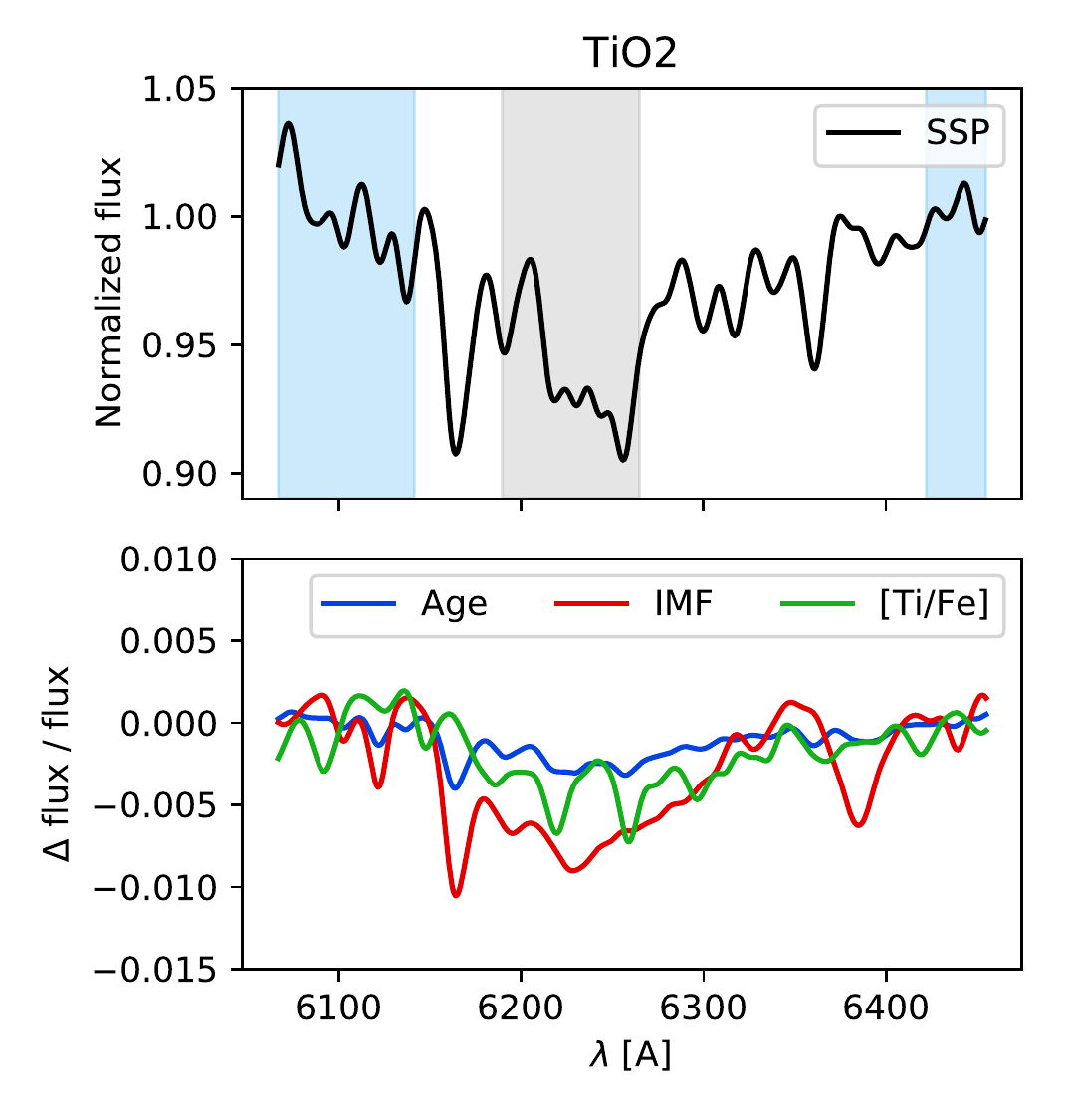}
\caption{The full-index-fitting (FIF) approach. The top panels show the Mgb\,5177 (left) and the TiO$_2$ (right) spectral features, normalized using the index pseudo-continua (blue shaded regions). In the FIF approach, every pixel within the central bandpass (grey area) is fitted to measure the stellar population parameters. The black line correspond to a model of solar metallicity [M/H]=0, [Mg/Fe]=0, [Ti/Fe]=0, and Kroupa-like IMF, at a resolution of 200 \kms. The FIF approach breaks the degeneracies more efficiently than the standard line-strength analysis since every pixel responds differently to changes in the stellar population properties. Colors in the bottom panels show the relative change in the spectrum after varying different stellar population parameters. For reference, the IMF slope was varied by $\Delta \Gamma_\mathrm{B}=1$, the metallicity and abundance ratios by 0.2 dex, and the age by 2 Gyr.}
\label{fig:indices}
\end{figure*}

Measuring detailed stellar population properties, and in particular IMF variations, from the absorption spectra of unresolved stellar populations requires a precise and reliable comparison between stellar population models and data. Two main approaches are usually followed. The standard line-strength analysis makes use of the equivalent width of well-defined spectral features to derive the stellar population properties of a given spectrum \citep[e.g.][]{Burstein84}. The main advantage of this method is that it focuses on relatively narrow spectral regions where the bulk of the information about the stellar population properties is encoded. Moreover, these narrow spectral features have been thoroughly studied and their dependence on the different stellar population parameters has been extremely well characterized \citep[e.g.][]{Worthey94,cat,TMB:03,Schiavon07,johansson12}. However usually only a handful of these indices can be analysed simultaneously, which may lead to degeneracies in the recovered stellar population properties \citep{Pat11}.

Additionally, and thanks to the development of intermediate resolution stellar population models, full spectral fitting techniques are now widely adopted. \citep[e.g.][]{CF05,Ocvirk06,Conroy09,Cappellari17,Wilkinson17}. Instead of focusing on specific absorption features, this second approach aims to fit every pixel across a relatively wide wavelength range ($\sim1000$\AA). Since every pixel is treated as an independent measurement, S/N requirements are lower than for a line-strength analysis, and with a given S/N ratio, degeneracies in the recovered stellar population parameters tend to be weaker using full spectral fitting algorithms \citep{Pat11}. Despite presenting clear advantages over the use of line-strength indices, the number of free parameters is also higher with full spectral stellar population fitting \citep[e.g.][]{Conroy18}, increasing the computational cost. Moreover, since every pixel in the spectrum is treated equally, the information about the stellar population properties, concentrated in narrow features, might get diluted \citep{labarbera}. 

A hybrid approach, in between line-strength analysis and full spectral fitting, is also possible. The idea consists of selecting key absorption features, where the information about the stellar population properties is concentrated. Then, instead of calculating the equivalent widths, every pixel within the feature is fitted after normalizing the continuum using the index definition \citep{MN15d}. In practice, this is a generalization of the line-strength analysis: while in the standard approach the equivalent width is measured with respect to a well-defined continuum, the hybrid method quantifies the depth of each pixel with respect to the same continuum definition. Fig.~\ref{fig:indices} exemplifies normalized Mgb\,5177 and TiO$_2$ spectral features.

This hybrid approach, or full index fitting (FIF), presents significant advantages. First, only specific spectral features are fitted, where the information is concentrated and the behavior of the stellar population properties is well determined. This allows for a low number of free parameters, as in the standard line-strength analysis, reducing the computational time. This is a key feature given the large number of spatial bins provided by the MUSE spectrograph. Moreover, since the continuum is fitted by a straight line using the index definition, the FIF approach is insensitive to large-scale flux calibration issues in the data. Compared with the line-strength analysis, the main advantage of the FIF method is that the number of independent observables is significantly increased. For example, in a standard index-index diagram (e.g. $H_\beta$ vs Mgb\,5177), only two measurements are compared to the model predictions. However, using FIF, the same two indices lead to more than $\sim100$  measurements (at the MILES resolution). In practice, adjacent pixels in the observed spectrum of a galaxy are correlated so the effective improvement of the S/N does not necessarily scales as $\sqrt{N_{pix}}$. In addition to these practical advantages, the use of FIF has a crucial characteristic: each pixel in a given spectral feature depends differently on the stellar populations parameters, as shown in Fig.~\ref{fig:indices}. This implies that not only the S/N requirements are lower but also that degeneracies among stellar population properties become weaker (see Fig.~\ref{fig:corner}).

\subsection{Application to F3D data} \label{sec:f3ddata}

In order to apply the FIF method to the F3D data of FCC167, we first measured the stellar kinematics (mean velocity and velocity dispersion) of each individual spatial bin (see \S~\ref{sec:data}) using the pPXF code presented in \citet{ppxf}. For consistency, we fed pPXF with the same set of MILES models used for the stellar population analysis. Spectral regions potentially contaminated by ionized gas emission were masked in this first step of the fitting process. Note that, for consistency, we did not use the kinematics described in the survey presentation paper \citep{f3d}, as we made different assumptions in the modeling process.

An important limitation of the MUSE spectrograph is the relatively {\it red} wavelength coverage, which starts at $\lambda=$4650\AA, and consequently, the only reliable age sensitive feature in the observed wavelength range is the H$_\beta$ line. However, H$_\beta$ is known to depend, not only on the age, but also on some other stellar population properties. In particular, H$_\beta$ shows a significant sensitivity to the [C/Fe] abundance ratio \citep{conroy,LB16}. Unfortunately, there are no prominent C-sensitive features in the MUSE data, making the use of H$_\beta$ unreliable. We overcome this problem by measuring the luminosity weighted age of FCC\,167 using the pPXF regularization scheme \citep{Cappellari17}, which can provide robust stellar population measurements \citep{McDermid15}. Given the sensitivity of recovered star formation histories on the assumed IMF slope \citep{FM13}, we regularized over the age--metallicity--IMF slope parameter space, and then we fixed the pPXF best-fitting age throughout the rest of the stellar population analysis.

Given the wavelength coverage of the [$\alpha$/Fe]-variable MILES models ($\lambda\lambda = 3540-7410$\AA), we based our stellar population analysis of FCC\,167 on features bluewards of $\lambda\sim7000$. This wavelength range shows a wide variety of stellar population-sensitive features and is much less affected by telluric absorption than the near-IR regime. To constrain the metallicity and [$\alpha$/Fe] ratio we focused on the Fe\,5270, Fe\,5335, and Mgb5177 indices \citep{trager}. Although all $\alpha$ elements are varied in lock-step in the MILES models, our only [$\alpha$/Fe]-sensitive feature is the Mgb5177 absorption feature. Hence, it is only the [Mg/Fe] abundance ratio which is constrained by our analysis. In the MUSE wavelength range, the most important IMF sensitive features are the aTiO \citep{Jorgensen94}, the TiO$_1$ and the TiO$_2$ absorptions \citep{serven}. We therefore also included the effect of [Ti/Fe] as an additional free parameter using the response functions of \citet{conroy}. The [Ti/Fe] is not treated in the same way as the [Mg/Fe] ratio, as the same response function is assumed irrespective of the value of the other stellar population parameters \citep[see][]{Spiniello15}. Effectively, [Ti/Fe] has a relatively mild effect on the selected features. Finally, because the effect of [C/Fe] in our set of features balances out that of the [$\alpha$/Fe] \citep{LB16}, and both ratios are expected to track each other \citep{johansson12}, we neglect the [$\alpha$/Fe] sensitivity of the MILES models beyond $\lambda=5400$\AA.

In short, we follow the FIF stellar population fitting approach described in \S~\ref{sec:fif} focusing on six spectral features (Fe\,5270, Fe\,5335, Mgb5177, aTiO, TiO$_1$, and TiO$_2$). We fit for four stellar population parameters, namely, metallicity, [Mg/Fe], [Ti/Fe] and IMF slope ($\Gamma_\mathrm{B}$). The age was fixed to that measured using pPXF, and the effect of the [Mg/Fe] was only considered for wavelengths $\lambda\leqq5400$\AA. The implications of these assumptions on the recovered stellar population parameters are shown and discussed in \S~\ref{sec:etg}.

\begin{figure}
\centering
\includegraphics[width=\hsize]{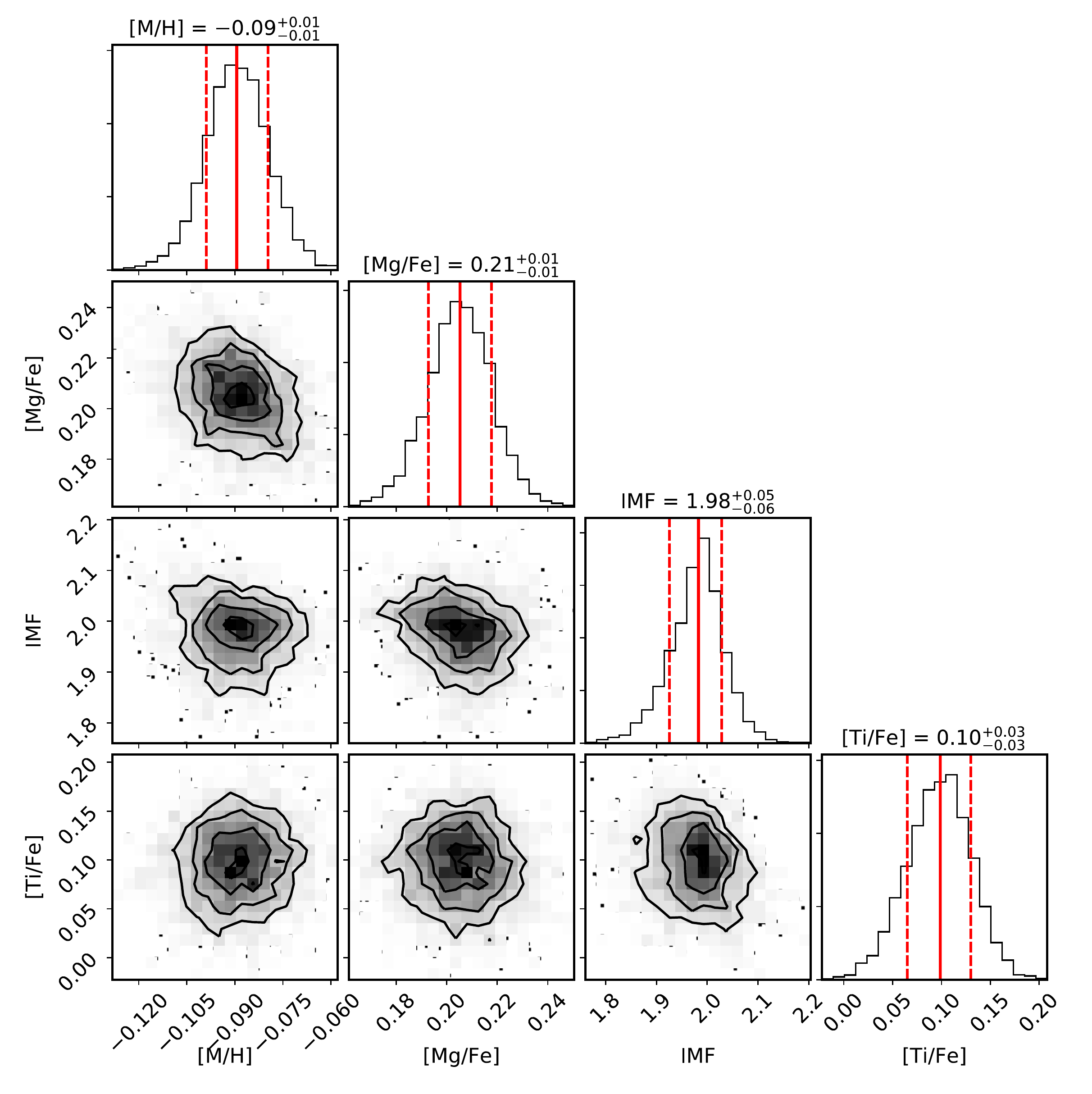}
\caption{Best-fitting stellar population properties. Posterior distributions for one of the spatial bins of FCC\,167 showing how our FIF approach is able to recover the stellar population properties with high precision, breaking the degeneracies among them. Vertical red lines indicate the best fitting value (solid line) and the 1$\sigma$ uncertainty (dashed lines). These values are quoted on top of the posterior distributions. The luminosity-weighted age of this bin is 12.7 Gyr.}
\label{fig:corner}
\end{figure}

In order to compare models and data, we implemented the same scheme as in \citet{MN18}. We used the {\it emcee} Bayesian Markov chain Monte Carlo sampler \citep{emcee}, powered by the Astropy project \citep{astropya,astropyb}, to maximize the following objective function

\begin{equation}~\label{eq:min}
 \ln ({\bf O} \,  | \, {\bf S} ) = -\frac{1}{2}  \mathlarger{\sum}_n \bigg[ \frac{(\mathrm{O}_n - \mathrm{M}_n)^2}{\sigma_n^2}-\ln \frac{1}{\sigma_n^2}\bigg]
\end{equation}

\noindent where {\bf S} = $\lbrace\mathrm{\Gamma_B,[M/H],[Mg/Fe],[Ti/Fe]}\rbrace$. The summation extends over all the pixels within the band-passes of the selected spectral features. O$_n$ and M$_n$ are the observed and the model flux\footnote{This model flux is obtained by linearly interpolating a grid of MILES models.} of the $n$th-pixel, and $\sigma_n$ the measured uncertainty. Eq.~\ref{eq:min} is therefore just a Gaussian likelihood function, where the distance (scaled by the expected uncertainity) between data and models is minimized. For FCC\,167, model predictions M$_n$ were calculated at a common resolution of 250 \kms \ so they had to be calculated only once and not for every spatial bin. This resolution corresponds to the lowest measured in FCC\,167. Before comparing models and data, every MUSE spectrum was smoothed to match the 250 \kms resolution of the models using the velocity dispersion measurements from pPXF. Fig.~\ref{fig:corner} shows how our approach is able to break the degeneracies among the different stellar population parameters when applied to FCC\,167 F3D data.

Thanks to the Bayesian fitting scheme described above, we could further improve the robustness of our results by imposing informative priors. In particular, we used the best-fitting metallicity and IMF values from the regularized pPXF fit as gaussian priors on the final solution. This final improvement does not bias the solution, but it increases the stability of the recovered stellar population parameters as shown in Fig.~\ref{fig:sdss}. We assumed flat priors for the other free parameters (e.g. [Mg/Fe] and [Ti/Fe]).

\subsection{ETG scaling relations}~\label{sec:etg}

\begin{figure}
\centering
\includegraphics[width=8.9cm]{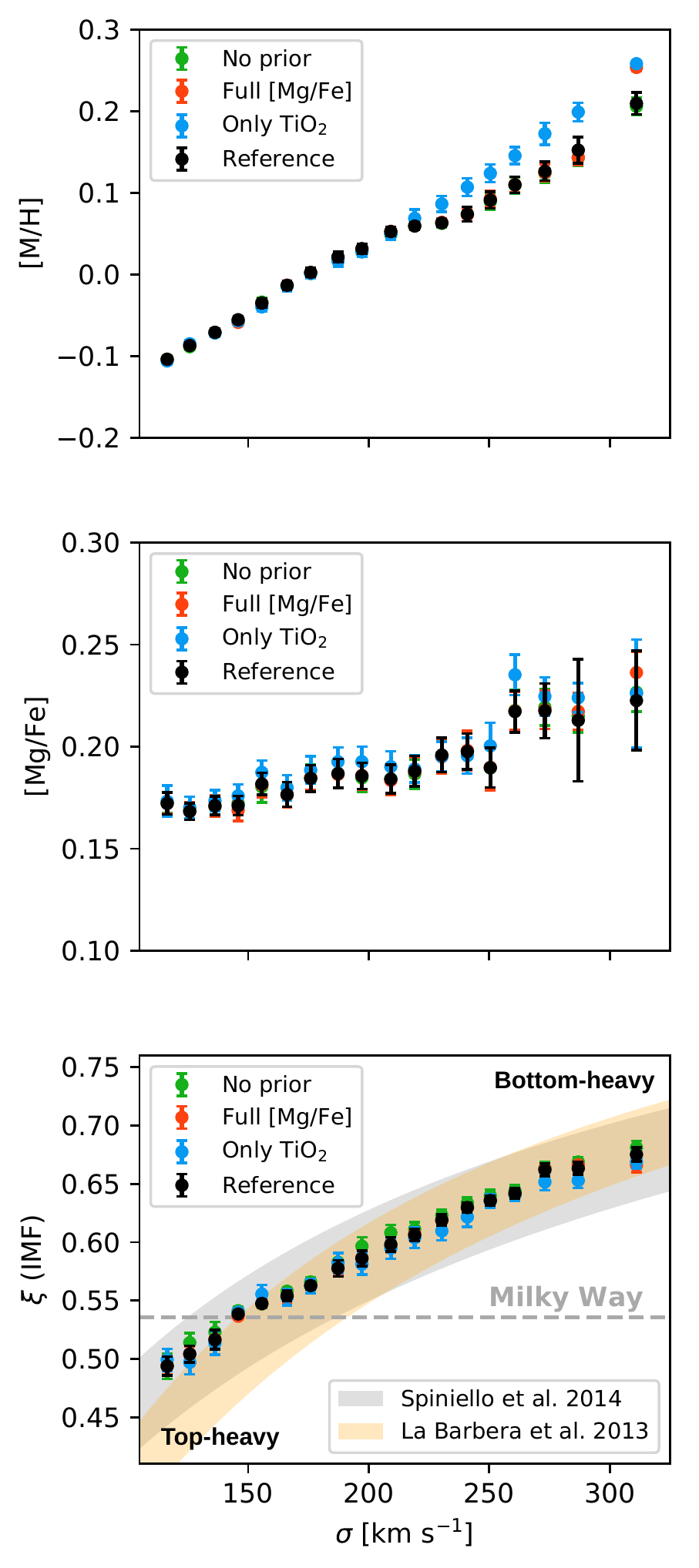}
\caption{From top to bottom, this figure shows the metallicity, [Mg/Fe], and IMF ($\xi$) scaling relations with galaxy velocity dispersion. The recovered trends agree with expectations from previous studies. In particular, the measured IMF-$\sigma$ relation is remarkably close to the results of \citet{labarbera} and \citet{Spiniello2013}, even though they rely on different model assumptions. Moreover, the use of $\xi$ as a proxy for the low-mass end IMF slope is able to unify different IMF parametrizations. The main difference between the trends shown in this figure and previous works is the fact that we measure a flatter trend between [Mg/Fe] and galaxy velocity dispersion. This is however, not related to the FIF fitting, but due to a combination of the metallicity-dependent [Mg/Fe] effect on the MILES models, plus the sensitivity of the Mgb\,5177 line to changes in the IMF slope (see Fig.~\ref{fig:indices}).}
\label{fig:sdss}
\end{figure}

\begin{figure*}
\centering
\includegraphics[width=9cm]{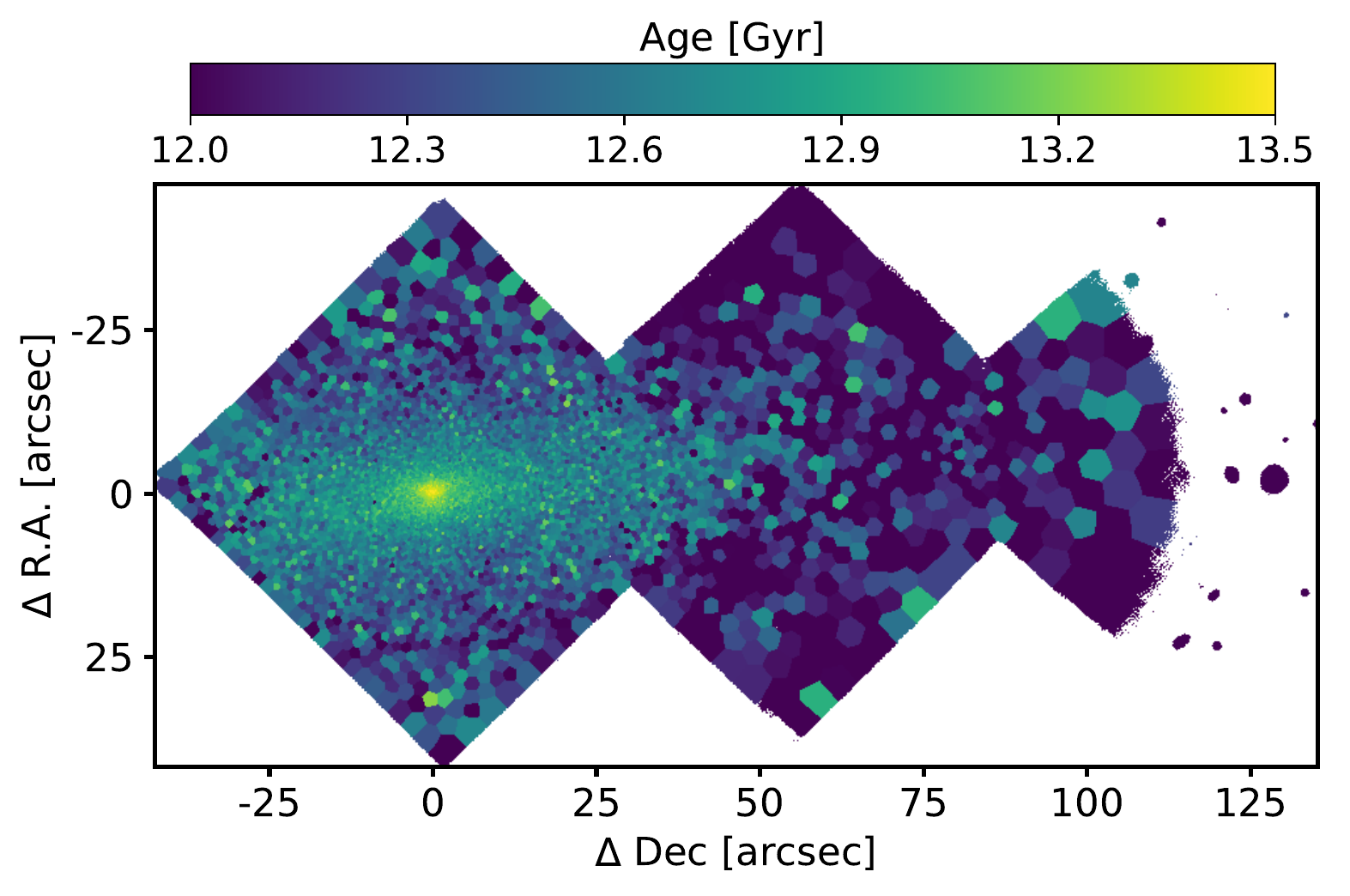}
\includegraphics[width=9cm]{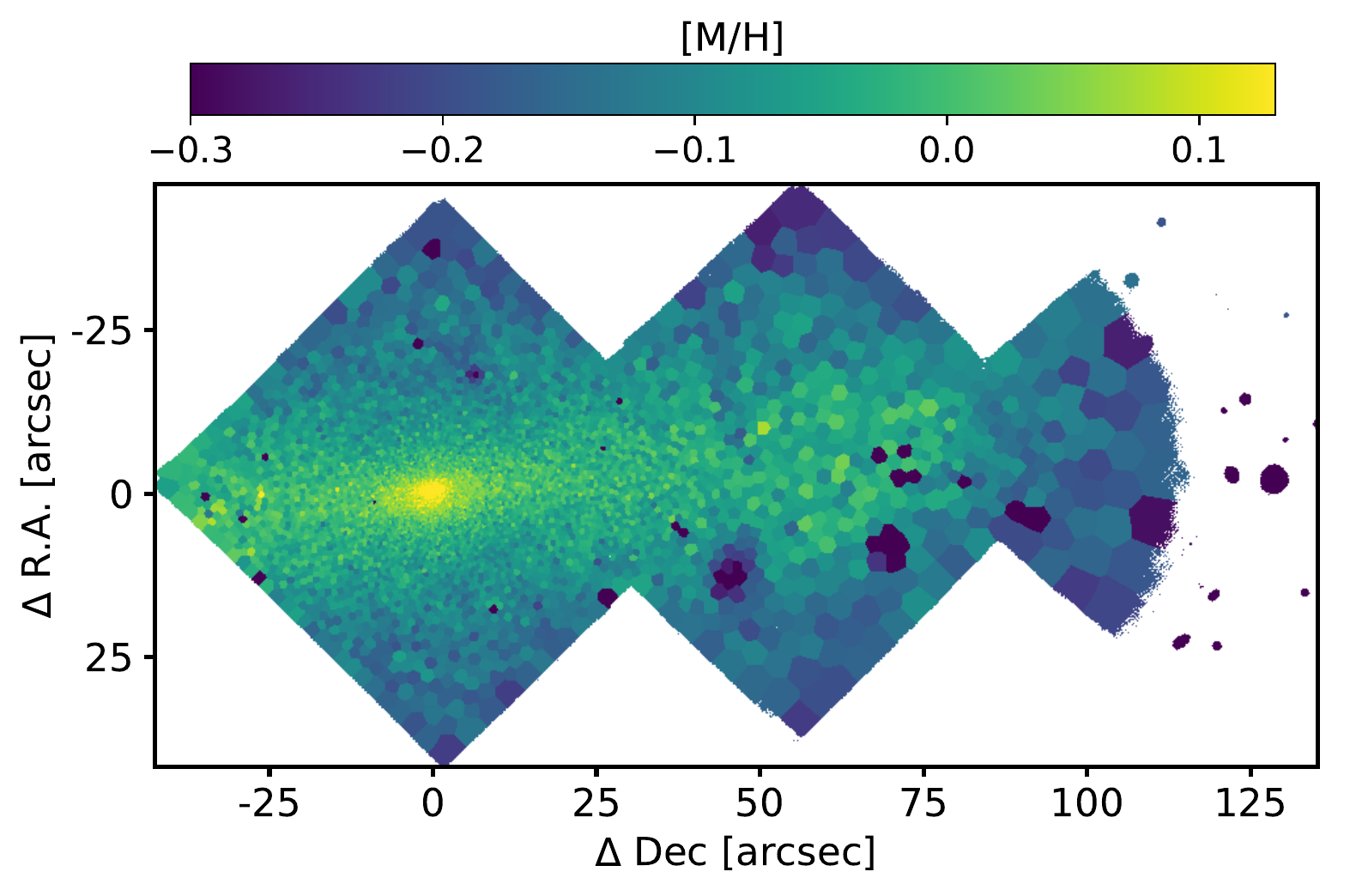}
\includegraphics[width=9cm]{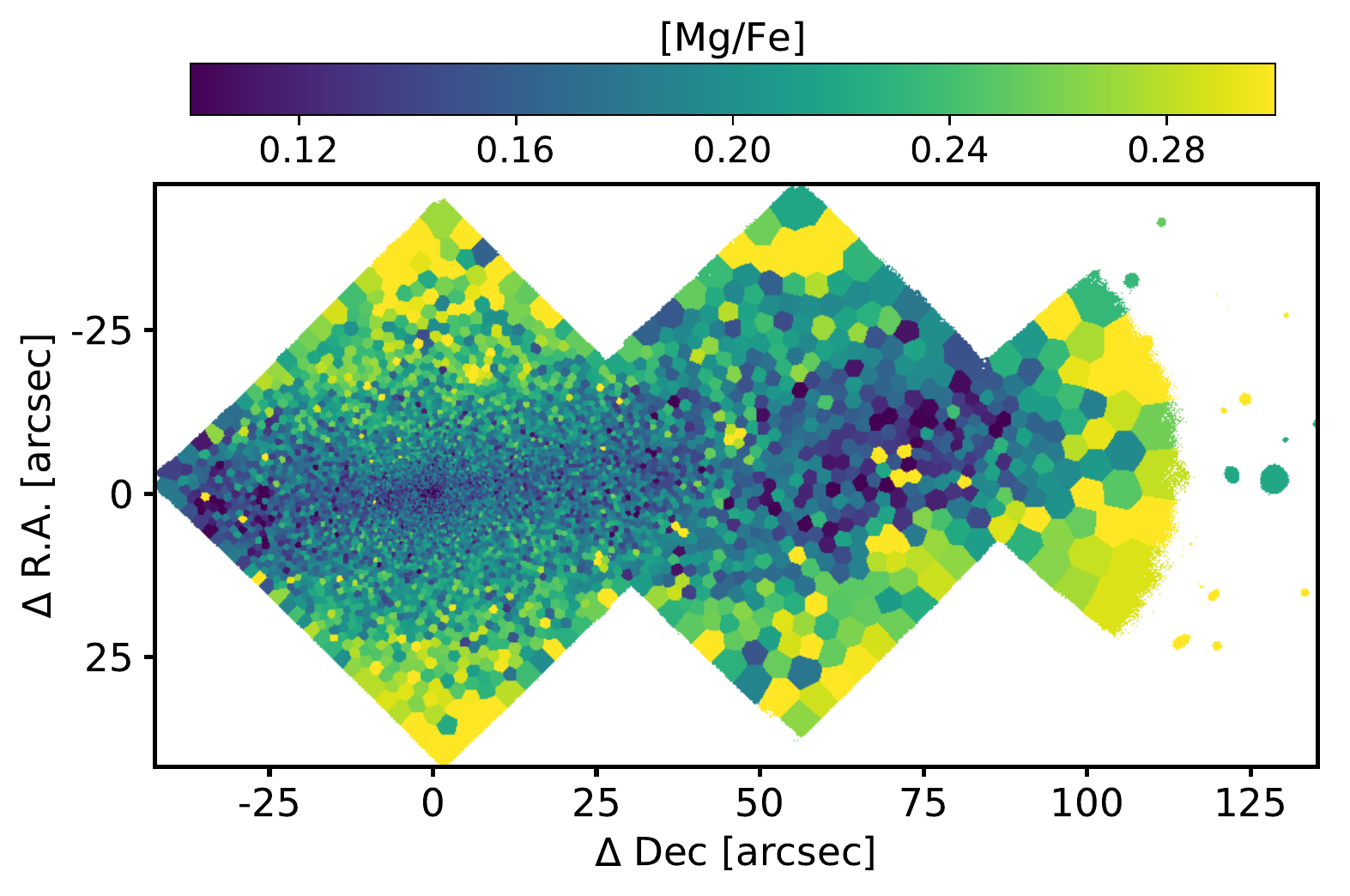}
\includegraphics[width=9cm]{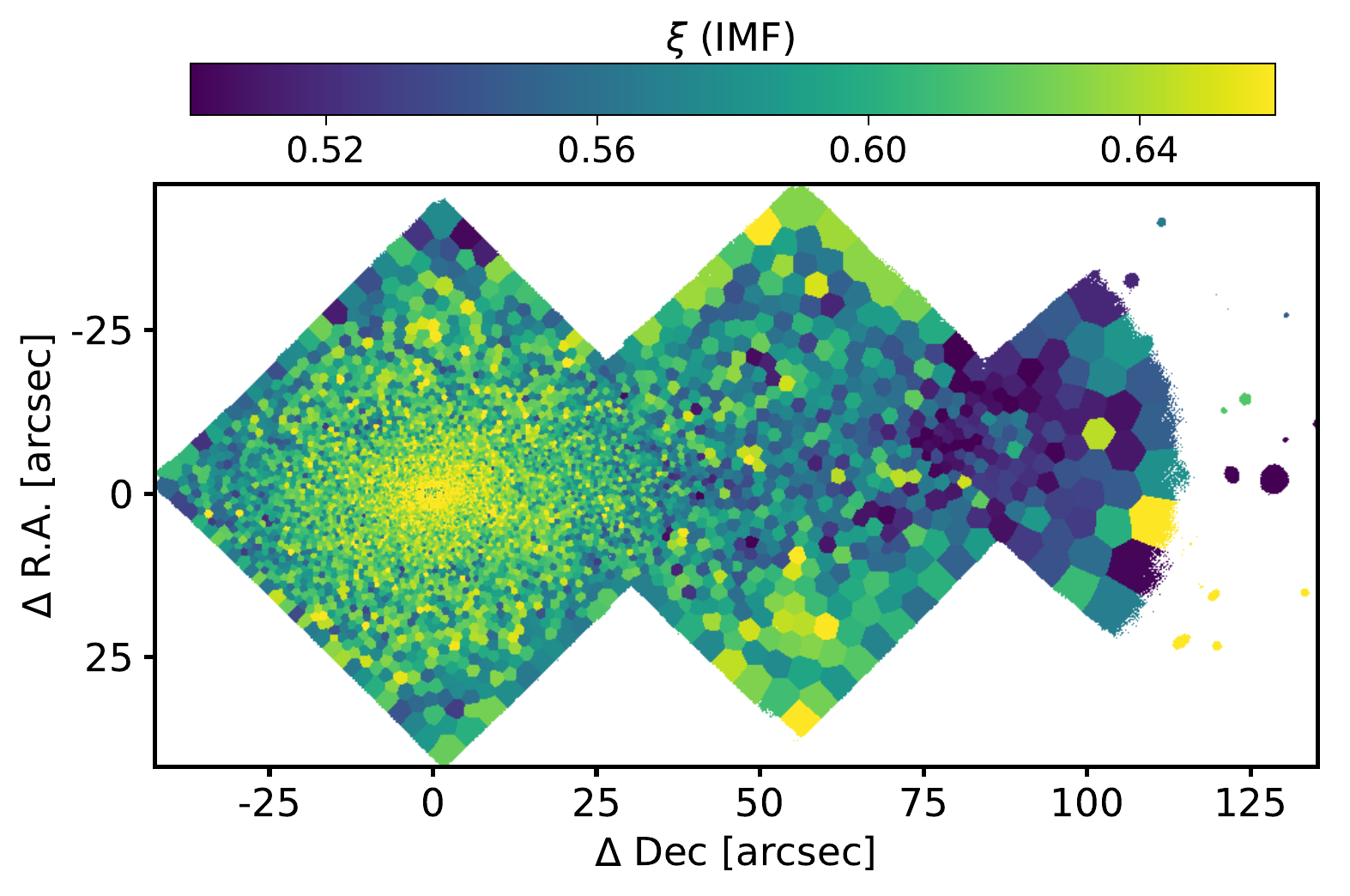}
\caption{Stellar population maps of FCC\,167. The top left panel shows the age map, as measured from pPXF, where the presence of a relatively older central component is clear. The metallicity (top right) and the [Mg/Fe] (bottom left) maps show the clear signature of a chemically evolved disk, confined within a vertical height of $\sim10$ arcsec, coinciding with the kinematically cold component observed in this galaxy \citep{f3d}. The IMF map on the bottom right exhibits however a different two-dimensional structure, not following the metallicity variations, but closely following the hot+warm orbits of this galaxy \citep{f3d}.}
\label{fig:maps}
\end{figure*}

\noindent The stellar population properties of ETGs follow tight scaling relations with galaxy mass \citep[e.g.][]{Worthey92,Thomas05,Kuntschner10}. Therefore, before attempting the stellar population analysis of FCC\,167, we test whether our FIF approach is able to recover these well-known trends. We made use of the ETG stacked spectra of \citet{labarbera}, based on the public Sloan Digital Sky Survey DR6 \citep{DR6}, and we fit them with our reference set up described above. In addition, to understand the effect of the different model assumptions on the recovered stellar population properties, we also performed a series of tests where we varied our fitting scheme. The first variation consisted of removing the informative priors. Second, we also tested the effect of allowing for the [Mg/Fe] variations to affect wavelengths beyond $\lambda=5400$\AA. Third, the robustness of the FIF approach, and its dependence on the set of indices was put to the test by removing all IMF sensitive features but TiO$_2$ from the analysis. The results of these tests are shown in Fig.~\ref{fig:sdss}.

The top panel of Fig.~\ref{fig:sdss} shows the metallicity--stellar velocity dispersion using our FIF approach. The expected trend, with galaxies with higher velocity dispersion (more massive) being more metal rich is recovered \citep[e.g.][]{Thomas10}. Only the case where the TiO$_2$ is our only IMF sensitive feature departs from the rest, proving that metallicity is well constrained by our approach.

The middle panel in Fig.~\ref{fig:sdss} shows how the [Mg/Fe] abundance ratio changes as a function of galaxy velocity dispersion. Again, as for the metallicity, all the different variations agree, and they show how more massive galaxies are more [Mg/Fe] enhanced. An important point should be noticed: the [Mg/Fe] -- velocity dispersion relation is flatter than expected \citep[e.g.][]{Thomas10,dlr11,LB14,Conroy14}. We checked that this flattening is not due to the FIF approach by repeating the analysis with the standard line-strength indices and finding the same result. The weak [Mg/Fe] trend with galaxy velocity dispersion results from the combination of two processes not explored in previous studies. First, the new MILES models consistently capture the effect of the [Mg/Fe], whose effect becomes weaker at sub-solar metallicities \citep{Vazdekis15}. This implies that for low-$\sigma$ galaxies, where metallicities are also low, a higher [Mg/Fe] is needed to match the data compared to other stellar population models. In consequence, low-$\sigma$ galaxies are found to have higher [Mg/Fe] than previously reported \citep{Aga17}. The second effect flattening the [Mg/Fe]-$\sigma$ relation has to do with the sensitivity of the Mgb\,5177 feature to the IMF, as shown for example in the left panels of Fig.~\ref{fig:indices}. A variation in the IMF slope of $\delta \Gamma\sim1$ as typically observed in massive ETGs leads to a change in the Mgb\,5177 index which is equivalent to an increment of $\delta$[Mg/Fe]$\sim0.1$. Hence, the depth of the Mgb\,5177 absorption feature, traditionally interpreted as a change in the [Mg/Fe] \citep{Thomas05}, is also driven by a steepening in the IMF slope of massive ETGs \citep{conroy}.

Finally, the bottom panel of Fig.~\ref{fig:sdss} shows the recovered trend between IMF slope ($\xi$) and galaxy mass, in very good agreement with previous works. Shaded regions indicate the trends found by \citet{labarbera} and \citet{Spiniello2013} with the typical 1$\sigma$ uncertainty. It is clear from this panel that with our FIF approach, we do not only recover the expected trends, but also with a smaller uncertainty. This is ultimately due to the larger number of pixels used in the analysis and to the effect of model systematics \footnote{\citet{labarbera} also included in their error budget uncertainties on the treatment of abundance patterns and the emission correction on Balmer lines.}. Even using a single IMF indicator, in this case the TiO$_2$ feature, we are able to robustly measure the IMF in ETGs. Moreover, Fig.~\ref{fig:sdss} also shows how our $\xi$ definition is able to unify IMF measurements based on different IMF parametrizations. It is also worth mentioning that our FIF approach, and the works of \citet{labarbera} and \citet{Spiniello2013} are based on different sets of indices, and even different stellar population model ingredients, making the agreement among the three approaches even more remarkable.

\section{Results} \label{sec:results}

Having validated the FIF approach with the \citet{labarbera} stacked spectra, we applied our stellar population fitting scheme to the F3D data cube of FCC\,167. As mentioned in \S~\ref{sec:data}, after spatially binning the three MUSE pointings of FCC\,167, we measured the stellar population properties in $\sim$6000 independent Voronoi bins, mapping the two-dimensional structure of the galaxy. The main results of our analysis are shown in Fig.~\ref{fig:maps}.

On the top left panel of Fig.~\ref{fig:maps}, the age map of FCC\,167 shows how this galaxy hosts old stellar populations at all radii, although slightly older in the center. These relatively older stars seem to track the bulge-like structure of FCC\,167 \citep[as shown in Figure 10 of][]{f3d}. Note that ages shown in Fig.~\ref{fig:maps} are luminosity-weighted values derived using pPXF (where IMF and metallicity were also left as free parameters in a 20 (age) x 10 ([M/H]) x 10 (IMF) model grid, covering the same parameter space as described in \S~\ref{sec:model}).

The metallicity (top right) and [Mg/Fe] (bottom left) maps show clear evidence of the presence of a chemically evolved thin disk. Interestingly, the age map appears partially decoupled from the chemical properties of FCC\,167. It is particularly striking how some of the oldest regions, where star formation ceased very early in the evolution of FCC\,167, show at the same time chemically evolved (e.g. metal rich and [Mg/Fe] poor) stellar populations.

Finally, the bottom right panel in Fig.~\ref{fig:maps} shows, for the first time, the IMF map of an ETG. The overall behavior of the IMF is similar to what was found previously in massive ETGs, namely, it is  only in the central regions where the fraction of low-mass stars appears enhanced with respect to the Milky Way expectations \citep[e.g.][]{MN15b,LB16,vdk17}. At a distance of $\sim$100 arcsec from the center of FCC\,167, a Milky Way-like IMF slope is found. However, an important difference with respect to previous studies should be noticed. The two-dimensional structure of the IMF in this galaxy is clearly decoupled from the chemical one, in particular, from the metallicity map \citep{MN15c}. Instead of following a disky structure, the IMF map of FCC\,167 appears much less elongated.

\section{Discussion} \label{sec:discussion}

The ground-breaking potential of the F3D project to understand the formation and evolution of ETGs through their stellar population properties is clear from the two-dimensional maps shown in Fig.~\ref{fig:maps}. In spatially unresolved studies, the global properties of ETGs appear highly coupled, as more massive galaxies are also denser, more metal-rich, more [Mg/Fe]-enhanced, older and with steeper (low-mass end) IMF slopes \citep[e.g.][]{Thomas10,labarbera}. In the same way, by collapsing the information of a galaxy into a one-dimensional radial gradient, valuable information about the stellar population parameters is lost. For example, in FCC\,167, both IMF slope and metallicity decrease smoothly with radius, but it is only when studying its full 2D structure that their different behavior becomes evident.

\subsection{The age - [Mg/Fe] discrepancy}
The [Mg/Fe] map in FCC\,167 is clearly anti-correlated with the age one (Fig.~\ref{fig:maps}), transitioning from old and low [Mg/Fe] populations in the center towards relatively younger and more Mg-enhanced outskirts. Under the standard interpretation, age and [Mg/Fe] should tightly track each other, as lower [Mg/Fe] is reached by longer star formation histories \citep[e.g.][]{Thomas99}. In FCC\,167, the age map and its chemical properties seem to describe two different formation histories. The [Mg/Fe] map, in agreement with the metallicity distribution, suggests that the outer parts of the galaxy formed rapidly, which lead to chemically immature stellar populations (high [Mg/Fe] and low metallicities). This picture for the formation of the outskirts of FCC\,167 is in agreement with the properties of the Milky Way halo \citep[e.g.][]{Venn04,Hayes18,Emma18} and other spiral galaxies \citep[e.g.][]{Vargas14,Molaeinezhad17}. The inner metal-rich, low [Mg/Fe] regions would have formed during a longer period of time, leaving enough time to recycle stellar ejecta into new generations of stars. However, this scenario would imply a relatively younger inner disk, which is not evident from the age map. This apparent contradiction suggests that, in order to truly understand the stellar population properties of ETGs, {\em SSP-glasses} are not enough, and a more complex chemical evolution modeling is needed, taking into account the time evolution of the different stellar population parameters and our limitations on the stellar population modeling side, in particular the coarse time-resolution inherent to old stellar populations. This apparent tension between age and chemical composition properties is not unique to FCC\,167 and has been reported in previous IFU-based studies \citep[e.g.][]{MN18}.

\subsection{The IMF - metallicity relation}
\begin{figure}
\centering
\includegraphics[width=\hsize]{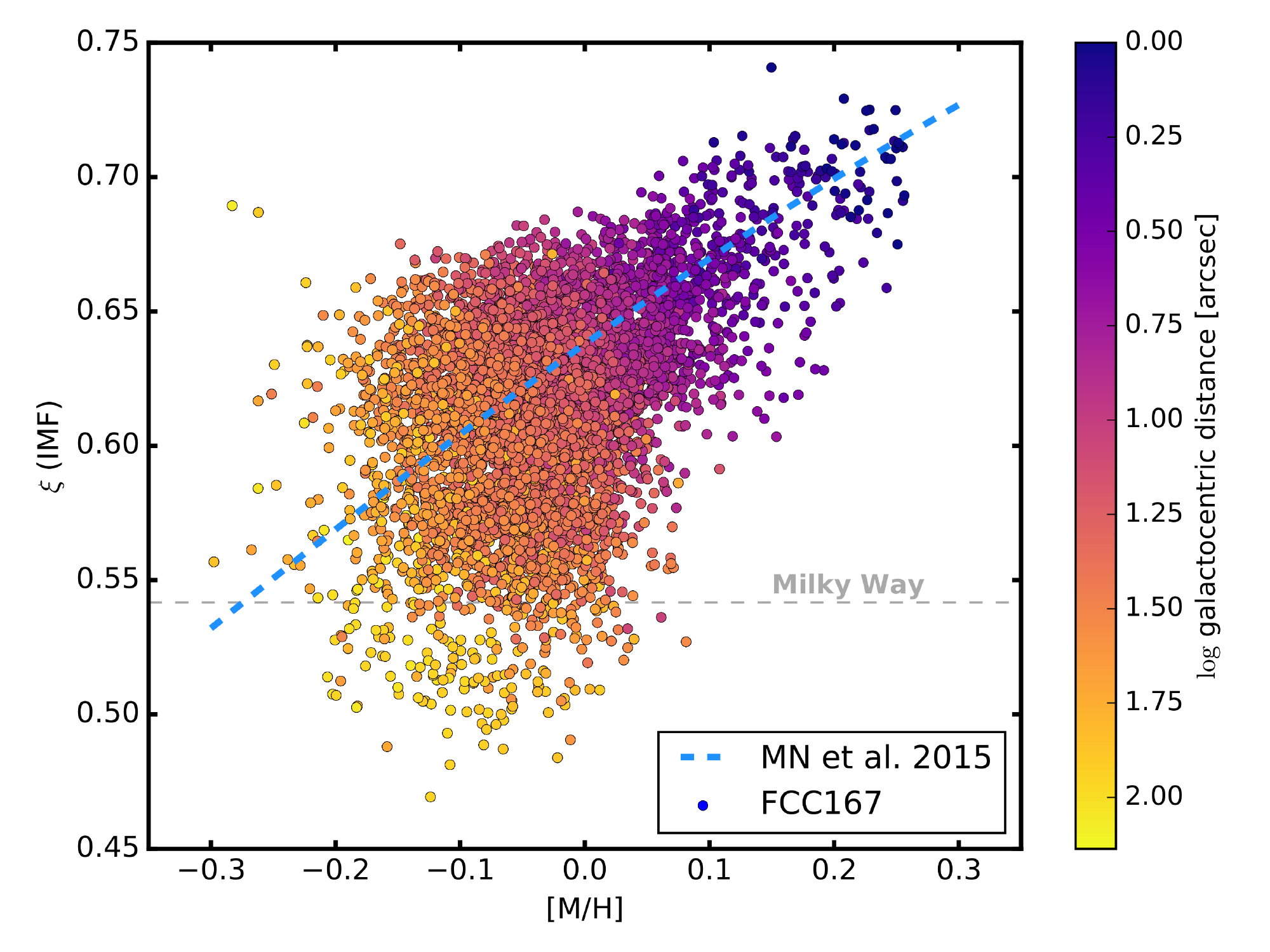}
\caption{IMF -- metallicity relation. The individual bins of FCC\,167 are shown color-coded by their distance to the center of the galaxy, compared with the empirical relation of \citet{MN15c}, shown as an blue dashed line. This relation agrees with the FCC\,167 measurements in the central regions of the galaxy (top right corner), but it does not hold for the outskirts, where an additional parameter is needed to explain the observed variations in the IMF. The $\xi$ ratio of the Milky Way \citep{mw} is shown as a grey dashed line.}
\label{fig:met}
\end{figure}

The complexity of understanding the stellar population properties of ETGs with the advent of IFU spectroscopy is further increased by the observed IMF variations. In Fig.~\ref{fig:met} we show how IMF and metallicity measurements compare in FCC\,167, where the dashed line shows the relation found by \citet{MN15c}. The agreement between the FCC\,167 measurements and the empirical IMF-metallicity is remarkable given all the differences in the stellar population modeling between the two studies, further supporting the robustness of our approach. However, it is clear that the IMF-metallicity relation of \citet{MN15c} is not enough to explain the 2D stellar population structure of FCC\,167. The core of FCC\,167 agrees with the expectations, but it clearly departs in the outer (lower metallicity and $\xi$) regions. This is not surprising given the fact that the measurements of \citet{MN15c} are biased towards the central regions of their sample of ETGs from the CALIFA survey \citep{califa}. It is worth mentioning that \citet{Sarzi18} found a good agreement between the metallicity and the IMF gradients in M\,87, tightly following the IMF-metallicity found by \citet{MN15c}. The decoupling in FCC\,167 is likely due to the fact that its internal structure has been preserved over cosmic time due to the lack of major disruptive merger events, which would have washed out the observed differences. Hence, massive lenticular galaxies appear as ideal laboratories to study the origin of the observed IMF variations.

\begin{figure*}
\centering
\includegraphics[width=14cm]{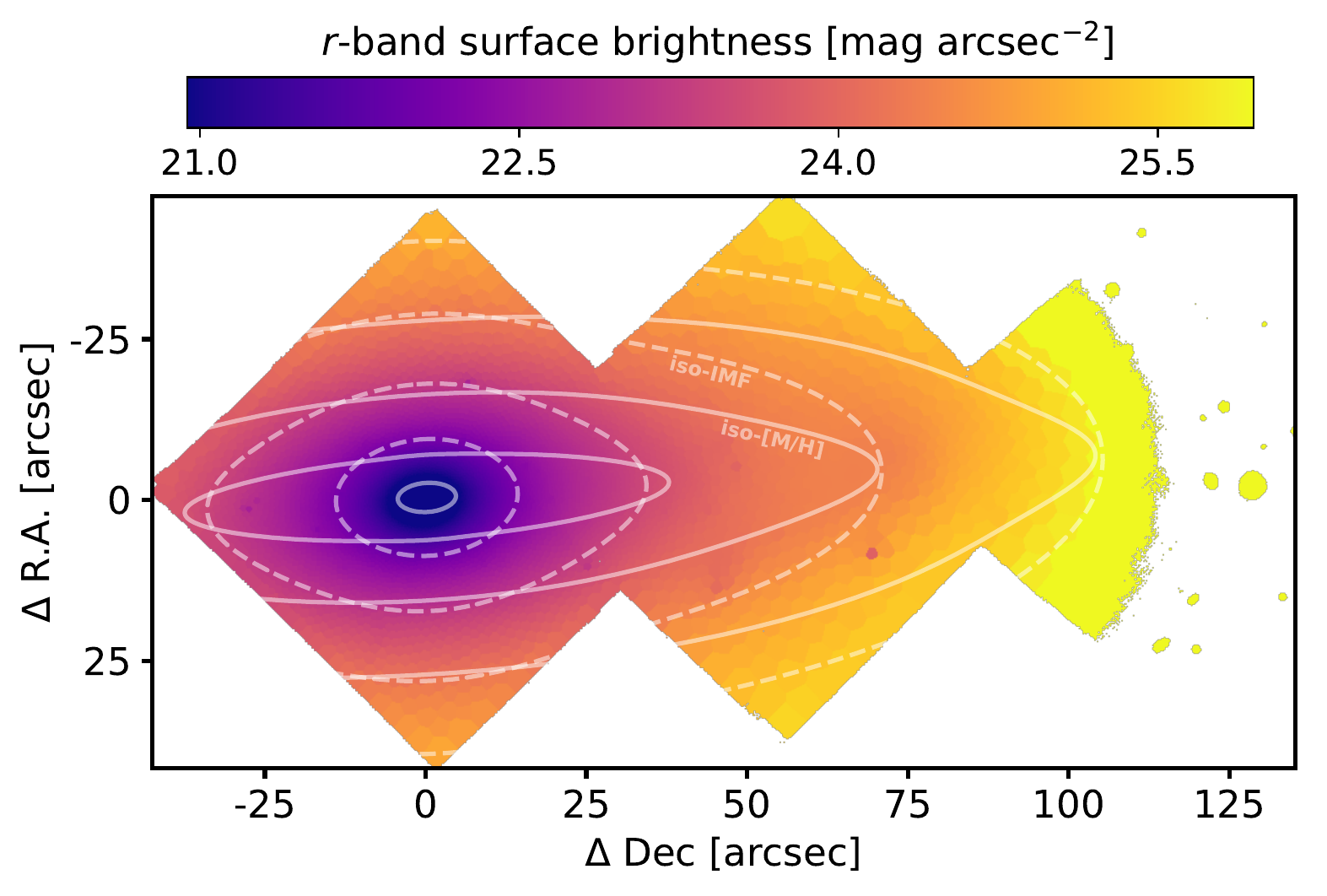}
\caption{Iso-metallicity vs iso-IMF contours. The surface brightness map of FCC\,167 is shown, as measured from the F3D data cube, with the iso-metallicity (solid lines) and iso-IMF contours (dashed lines) over-plotted. As in Fig.~\ref{fig:maps}, this figure shows how the two-dimensional IMF map does not exactly follow the metallicity variations. IMF variations appear more closely related to the surface brightness distribution, in particular around the central bulge. The metallicity distribution on the contrary is structured in a more disk-like configuration.}
\label{fig:iso}
\end{figure*}

The different behavior of the stellar population properties in FCC\,167 can also be seen by comparing the iso-metallicity and the iso-IMF contours. Fig.~\ref{fig:iso} shows the $r$-band surface brightness map of FCC\,167, as measured from the MUSE F3D datacubes. On top of the surface brightness map, the iso-metallicity and the iso-IMF contours are also shown. To  generate the contours in Fig.~\ref{fig:iso}, we fitted the stellar population maps with a multi-Gaussian-expansion model \citep{Emsellem94,Cappellari02}, as generally done with photometric data. This allows a smooth modeling of the large-scale behavior of the stellar population maps, which can then be easily transformed into iso contours.

The decoupling in the two-dimensional structure of the IMF and metallicity maps appears clearly in Fig.~\ref{fig:iso}. As described above, the metallicity  distribution follows a diskier structure, which is consistent with a long-lasting chemical recycling within the cold kinematic component of FCC\,167. The IMF on the other hand follows a rounder, more symmetric distribution.

\subsection{Stellar populations properties vs orbital decomposition}

\begin{figure*}
\centering
\includegraphics[width=9cm]{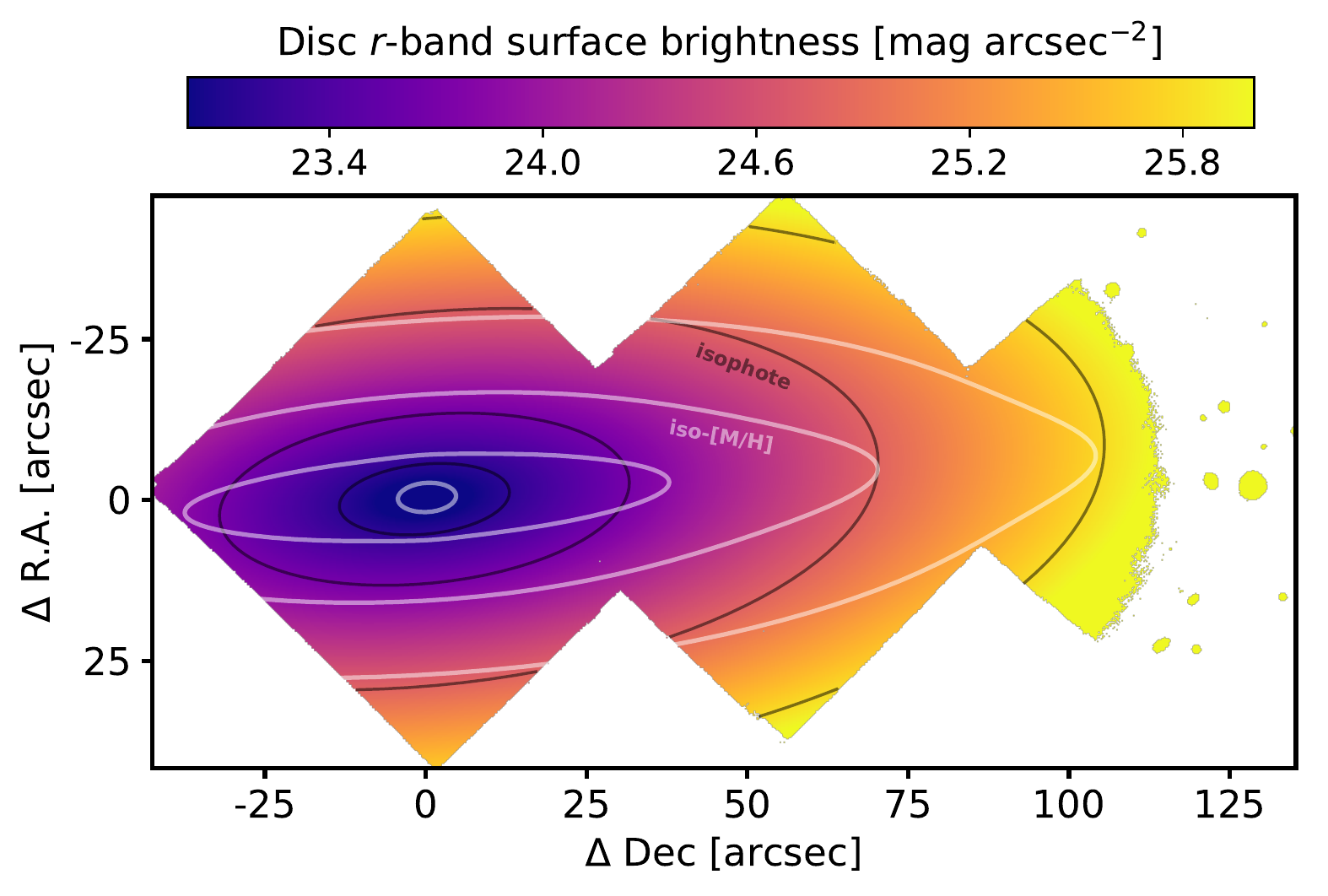}
\includegraphics[width=9cm]{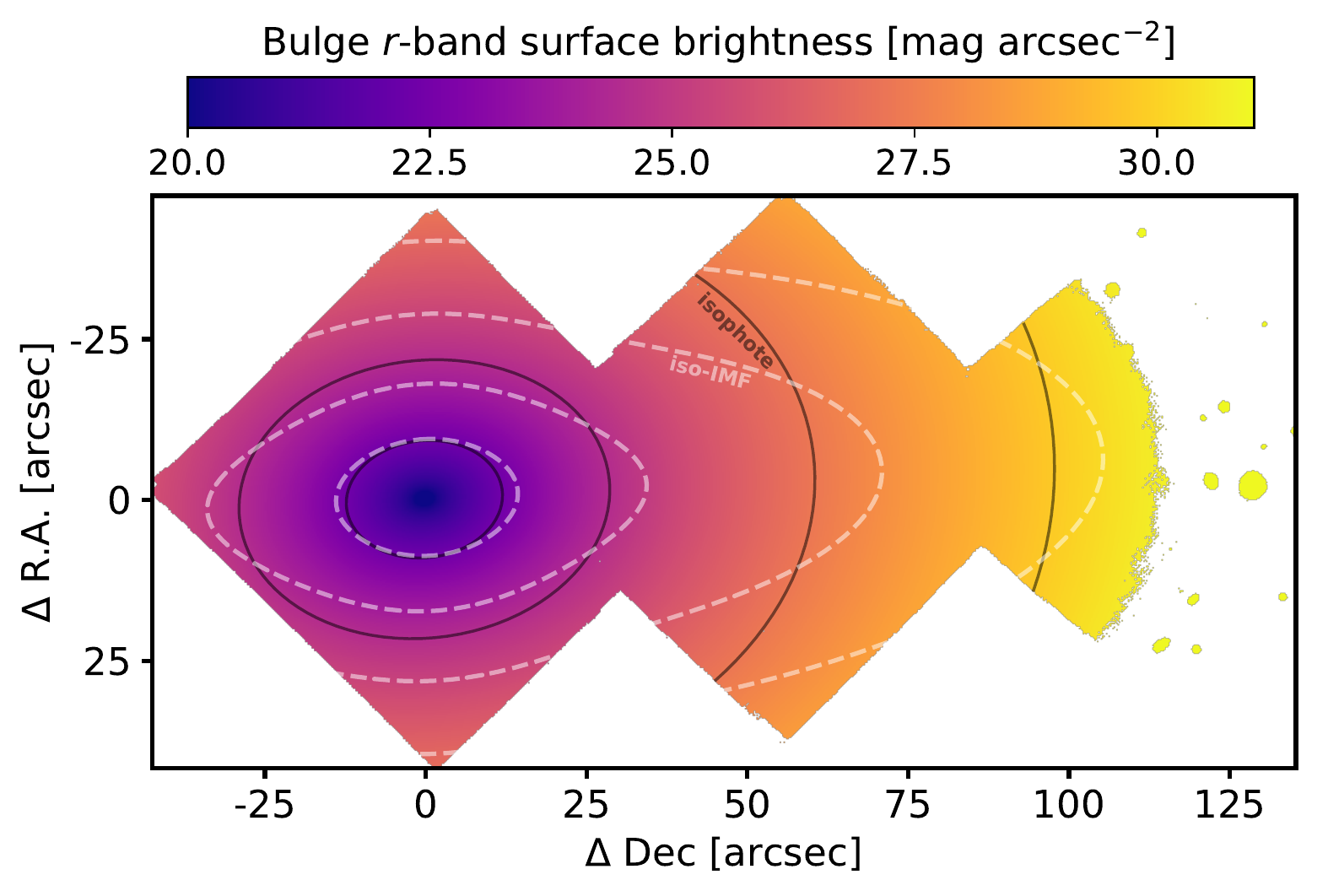}
\includegraphics[width=9cm]{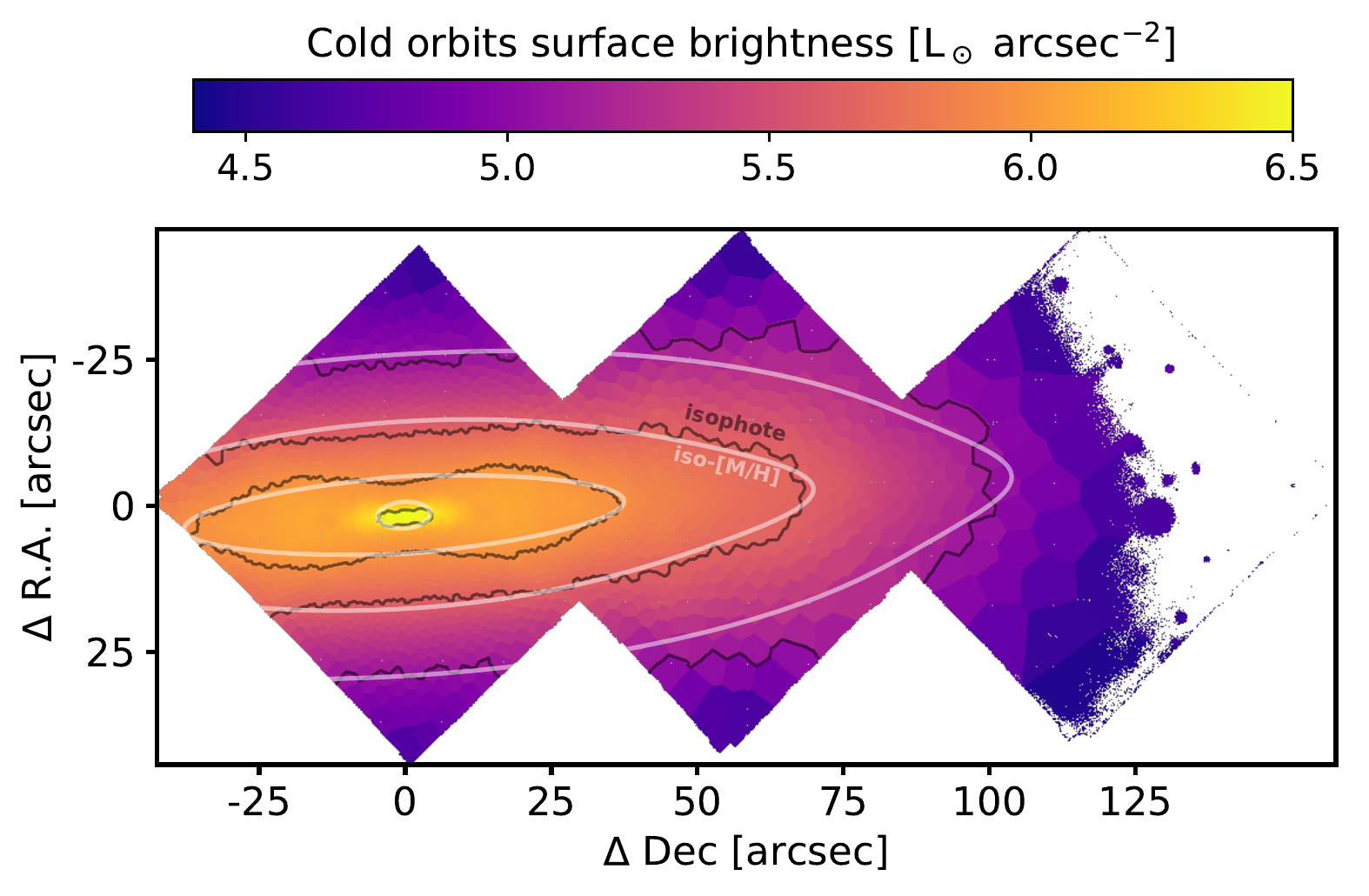}
\includegraphics[width=9cm]{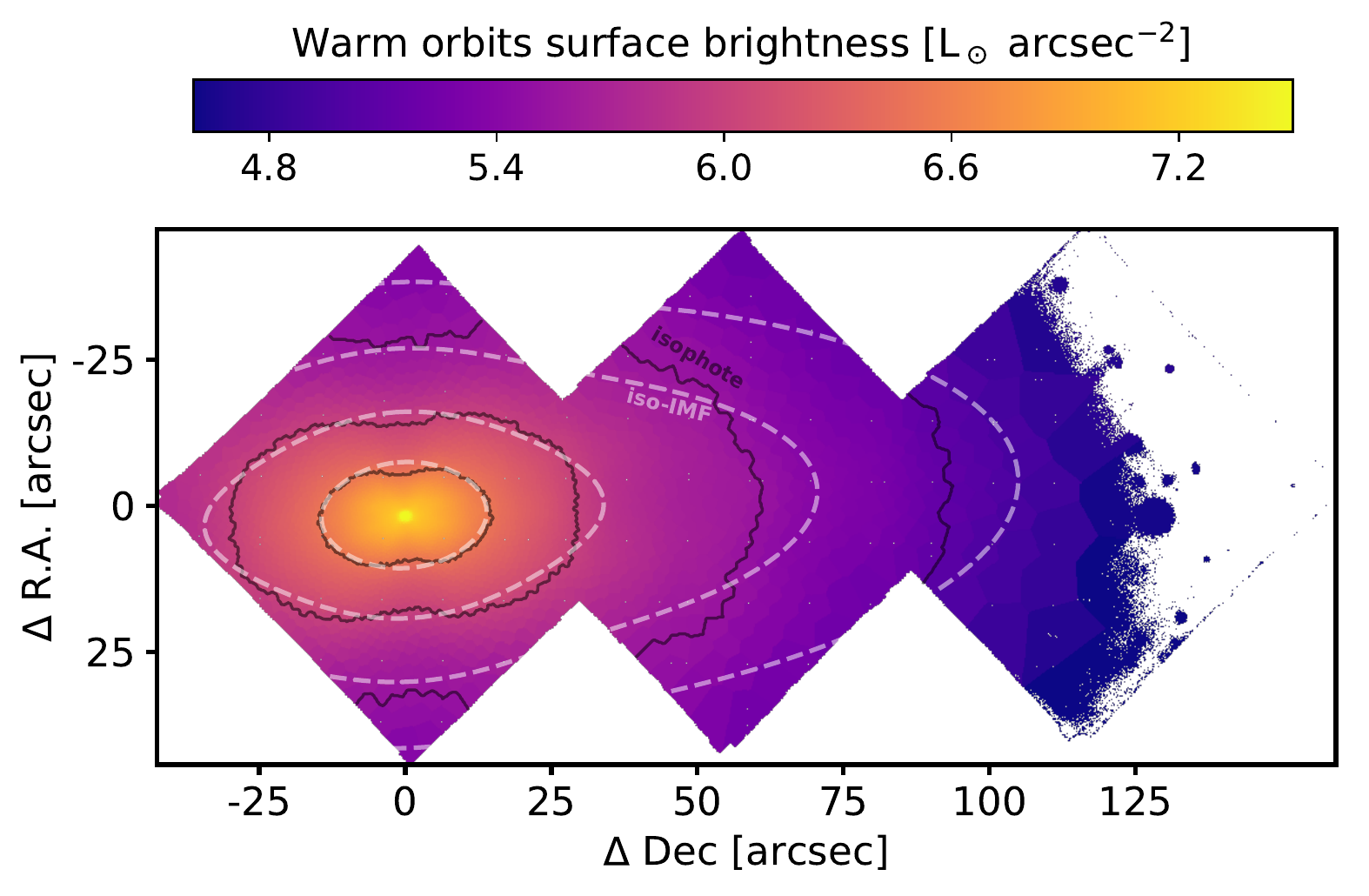}
\includegraphics[width=9cm]{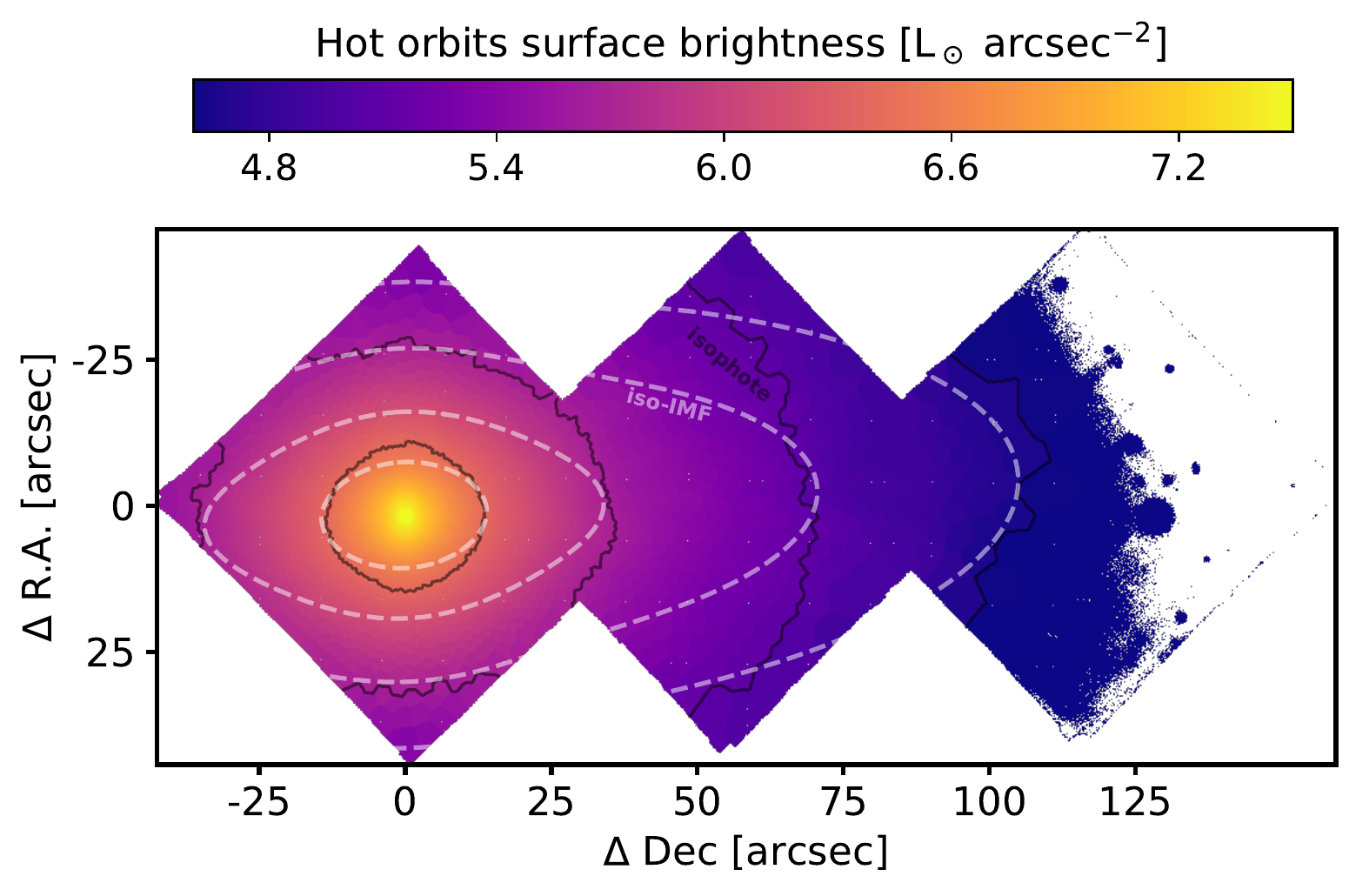}

\caption{Internal structure vs stellar population maps. Top panels show the comparison between metallicity and photometric disc (top left) and IMF and bulge (top right). Neither the disc nor the bulge seem to follow the structure of the stellar population maps. The three bottom panels compare the orbital decomposition of FCC\,167 with the stellar population properties. On the left, the agreement between the metallicity structure (white solid lines) and the spatial distribution of the cold orbits indicates that the chemically evolved (metal rich and [Mg/Fe] poor) structure observed in the stellar population maps corresponds to a cold stellar disc with an extended star formation history. IMF contours (white dashed lines) on the other hand closely follow the distribution of warm orbits (right), suggesting that the IMF was set at high $z$ during the assembly of the thick disc. Hot orbits on the contrary are decoupled from the stellar population properties, which might indicate that the bulge of FCC\,167 was formed through stellar heating processes. }
\label{fig:decomp}
\end{figure*}

In order to further investigate the connection between the internal structure of FCC\,167 and its stellar population properties we fit the synthetic $r$-band image with a bulge plus exponential disc model using {\it Imfit} \citep{Erwin15}. The top panels of Fig.~\ref{fig:decomp} show how the metallicity and IMF maps of FCC\,167 compare with its photometric decomposition. It is clear that the disc component does not capture the observed structure of the metallicity map. The agreement between IMF variations and bulge light distribution is slightly better, although the latter is much rounder. Thus, a simple bulge plus disc analysis is not able to capture the variation observed in the stellar population properties.

\citet{f3d} presented the Schwarzschild orbit-based decomposition \citep{remco08,Ling18} of FCC\,167, where they roughly distinguished between three types of orbits: cold ($\lambda_z > 0.7$), warm ($0.2 < \lambda_z < 0.7$), and hot ($\lambda_z < 0.2$). The bottom three panels in Fig.~\ref{fig:decomp} present the comparison between light distributions of these three types of orbits and the metallicity and IMF maps of FCC\,167. The coupling between the metallicity (and therefore the [Mg/Fe] ratio) maps and the spatial distribution of cold orbits is remarkable. This further supports the idea that the elongated structure shown by the chemical properties of FCC\,167 is indeed tracking a dynamically cold disc. A relatively more extended star formation history in this disc would naturally explain the high-metallicities and low [Mg/Fe] ratios. 

IMF variations on the contrary seem to be closely tracking the distribution of warm orbits, particularly in the central regions of FCC\,167. This result is rather unexpected, as it has been extensively argued that IMF variations are associated with the extreme star formation conditions within the cores of massive ETGs \citep[e.g.][]{MN15a,MN15b,vdk17}. The comparison shown in Fig.~\ref{fig:decomp} suggests, however, that the IMF was set during the early formation of the warm (thick disc) component of FCC\,167, where the pressure and density conditions may have had an impact on the shape of the IMF \citep{Chabrier14,Jerabkova18}. The weak correlation between stellar population properties and hot orbits might have strong implications for our understanding of bulge formation, as it suggests that most of the stars belonging to this dynamically hot structure were not born hot, but they must have been heated up at a later stage \citep[e.g.][]{Grand16,GK18}. 

From Fig.~\ref{fig:decomp} it becomes clear that the orbital decomposition offers a more meaningful and physically motivated framework than a standard photometric analysis \citep{Zhu18}. The connection with the metallicity and IMF distribution opens an alternative approach to understand the origin of the stellar population radial variations, in particular, in lenticular galaxies as FCC\,167 with a rich internal structure. Note however than while the stellar population maps are integrated quantities, the orbital analysis shown in Fig.~\ref{fig:decomp} is a decomposition into different orbit types, and this has to be taken into account before any further interpretation. For example, the iso-IMF contours are more elongated than the isophotes of the warm in the outskirts of FCC\,167 because at large radii the flux starts to be dominated by cold orbits. A more quantitative comparison between stellar population properties and orbital distributions will be presented in an upcoming F3D paper.

\section{Summary and conclusions}\label{sec:summary}

We have shown that the spatially-resolved stellar population properties of ETGs can be robustly measured thanks to the high quality MUSE-IFU data from the Fornax 3D project, combined with a novel approach which benefits from the advantages of both line-strength analysis and full-spectral fitting techniques. The analysis tools described in this work have allowed us to measure the two-dimensional stellar population property maps of the massive, S0/a galaxy FCC\,167. 

The chemical properties (i.e. metallicity and [Mg/Fe]) show a clear disc-like structure, associated with the cold orbital component of FCC\,167. IMF variations roughly follow the radial metallicity variation, in agreement with previous studies, but with a clearly distinct spatial distribution. Iso-IMF contours are much rounder than the iso-metallicity ones, and seem to follow the distribution of hot and warm orbits in this galaxy. These results suggests that metallicity can not be the only driver of the observed IMF variations in ETGs.

The comparison between the orbital decomposition and the stellar population properties provides a physically meaningful framework that captures the underlying IMF and metallicity variations in FCC\,167 better than a standard photometric decomposition. Our analysis describes a scenario where the IMF was set during the early formation of the stars with relatively warm orbits. The formation of the cold orbital component took place during a more extended period of time, leading to metal-rich and [Mg/Fe]-poor stellar populations. We argue that the time difference between the assembly of these two components is too short to be measured in old stellar populations. The chemical properties of ETGs, regulated by the ejecta of massive stars, would therefore be a finer timer than SSP-equivalent ages. Finally, the hot orbital component of FCC\,167 appears decoupled from the stellar population properties, suggesting that the formation of the bulge was likely due to a stellar heating process. The orbital-based analysis appears therefore as an insightful probe of the relation between galaxy structure and the emergence of the stellar population properties, and it will be explored in a upcoming work (Mart\'in-Navarro et al., {\it in prep}).

With the complex IMF variations shown by this object, understanding the stellar population properties of ETGs in the IFU era requires a deep change in both our modeling and our analysis of this type of galaxies. Well-known scaling relations supporting our picture of galaxy formation and evolution might be partially biased by the lack of spatial resolution. Moreover, the stellar population properties and the IMF in these objects have likely evolved over time \citep[e.g.][]{weidner:13,DM18,Fontanot18}, and at $z\sim0$ we only see an integrated snapshot of their lives. The assumption of SSP-like stellar populations in ETGs is starting to break down under the pressure of wider and deeper spectroscopic data, and the on-going projects within the F3D project will contribute to this change of paradigm \citep[e.g.][]{Pinna19}. In an upcoming paper, we will present and discuss the IMF variations for the whole sample of F3D galaxies, covering a wide range in masses and star formation conditions. 

\begin{acknowledgement}

We would like to thank Bronwyn Reichardt Chu for her useful comments and for the fruitful discussions, and also the referee for a careful and efficient revision of the manuscript. IMN ad JFB acknowledge support from the AYA2016-77237-C3-1-P grant from the Spanish Ministry of Economy and Competitiveness (MINECO). IMN acknowledges support from the Marie Sk\l odowska-Curie Individual {\it SPanD} Fellowship 702607. GvdV acknowledges funding from the European Research Council (ERC) under the European Union's Horizon 2020 research and innovation programme under grant agreement No 724857 (Consolidator Grant ArcheoDyn).  E.M.C. acknowledges financial support from Padua University through grants DOR1699945/16, DOR1715817/17, DOR1885254/18, and BIRD164402/16 

\end{acknowledgement}

\bibliographystyle{aa}  
\bibliography{IMF} 

\begin{thebibliography}{134}
\expandafter\ifx\csname natexlab\endcsname\relax\def\natexlab#1{#1}\fi

\bibitem[{{Adelman-McCarthy} {et~al.}(2008){Adelman-McCarthy}, {Ag{\"u}eros},
  {Allam}, {Allende Prieto}, {Anderson}, {Anderson}, {Annis}, {Bahcall},
  {Bailer-Jones}, {Baldry}, {Barentine}, {Bassett}, {Becker}, {Beers}, {Bell},
  {Berlind}, {Bernardi}, {Blanton}, {Bochanski}, {Boroski}, {Brinchmann},
  {Brinkmann}, {Brunner}, {Budav{\'a}ri}, {Carliles}, {Carr}, {Castander},
  {Cinabro}, {Cool}, {Covey}, {Csabai}, {Cunha}, {Davenport}, {Dilday}, {Doi},
  {Eisenstein}, {Evans}, {Fan}, {Finkbeiner}, {Friedman}, {Frieman},
  {Fukugita}, {G{\"a}nsicke}, {Gates}, {Gillespie}, {Glazebrook}, {Gray},
  {Grebel}, {Gunn}, {Gurbani}, {Hall}, {Harding}, {Harvanek}, {Hawley},
  {Hayes}, {Heckman}, {Hendry}, {Hindsley}, {Hirata}, {Hogan}, {Hogg}, {Hyde},
  {Ichikawa}, {Ivezi{\'c}}, {Jester}, {Johnson}, {Jorgensen}, {Juri{\'c}},
  {Kent}, {Kessler}, {Kleinman}, {Knapp}, {Kron}, {Krzesinski}, {Kuropatkin},
  {Lamb}, {Lampeitl}, {Lebedeva}, {Lee}, {French Leger}, {L{\'e}pine}, {Lima},
  {Lin}, {Long}, {Loomis}, {Loveday}, {Lupton}, {Malanushenko}, {Malanushenko},
  {Mandelbaum}, {Margon}, {Marriner}, {Mart{\'{\i}}nez-Delgado}, {Matsubara},
  {McGehee}, {McKay}, {Meiksin}, {Morrison}, {Munn}, {Nakajima}, {Neilsen},
  {Newberg}, {Nichol}, {Nicinski}, {Nieto-Santisteban}, {Nitta}, {Okamura},
  {Owen}, {Oyaizu}, {Padmanabhan}, {Pan}, {Park}, {Peoples}, {Pier}, {Pope},
  {Purger}, {Raddick}, {Re Fiorentin}, {Richards}, {Richmond}, {Riess}, {Rix},
  {Rockosi}, {Sako}, {Schlegel}, {Schneider}, {Schreiber}, {Schwope}, {Seljak},
  {Sesar}, {Sheldon}, {Shimasaku}, {Sivarani}, {Allyn Smith}, {Snedden},
  {Steinmetz}, {Strauss}, {SubbaRao}, {Suto}, {Szalay}, {Szapudi}, {Szkody},
  {Tegmark}, {Thakar}, {Tremonti}, {Tucker}, {Uomoto}, {Vanden Berk},
  {Vandenberg}, {Vidrih}, {Vogeley}, {Voges}, {Vogt}, {Wadadekar}, {Weinberg},
  {West}, {White}, {Wilhite}, {Yanny}, {Yocum}, {York}, {Zehavi}, \&
  {Zucker}}]{DR6}
{Adelman-McCarthy}, J.~K., {Ag{\"u}eros}, M.~A., {Allam}, S.~S., {et~al.} 2008,
  \apjs, 175, 297

\bibitem[{{Alton} {et~al.}(2018){Alton}, {Smith}, \& {Lucey}}]{Alton18}
{Alton}, P.~D., {Smith}, R.~J., \& {Lucey}, J.~R. 2018, \mnras, 478, 4464

\bibitem[{{Astropy Collaboration} {et~al.}(2018){Astropy Collaboration},
  {Price-Whelan}, {Sip{\H o}cz}, {G{\"u}nther}, {Lim}, {Crawford}, {Conseil},
  {Shupe}, {Craig}, {Dencheva}, {Ginsburg}, {VanderPlas}, {Bradley},
  {P{\'e}rez-Su{\'a}rez}, {de Val-Borro}, {Aldcroft}, {Cruz}, {Robitaille},
  {Tollerud}, {Ardelean}, {Babej}, {Bach}, {Bachetti}, {Bakanov}, {Bamford},
  {Barentsen}, {Barmby}, {Baumbach}, {Berry}, {Biscani}, {Boquien}, {Bostroem},
  {Bouma}, {Brammer}, {Bray}, {Breytenbach}, {Buddelmeijer}, {Burke},
  {Calderone}, {Cano Rodr{\'{\i}}guez}, {Cara}, {Cardoso}, {Cheedella},
  {Copin}, {Corrales}, {Crichton}, {D'Avella}, {Deil}, {Depagne}, {Dietrich},
  {Donath}, {Droettboom}, {Earl}, {Erben}, {Fabbro}, {Ferreira}, {Finethy},
  {Fox}, {Garrison}, {Gibbons}, {Goldstein}, {Gommers}, {Greco}, {Greenfield},
  {Groener}, {Grollier}, {Hagen}, {Hirst}, {Homeier}, {Horton}, {Hosseinzadeh},
  {Hu}, {Hunkeler}, {Ivezi{\'c}}, {Jain}, {Jenness}, {Kanarek}, {Kendrew},
  {Kern}, {Kerzendorf}, {Khvalko}, {King}, {Kirkby}, {Kulkarni}, {Kumar},
  {Lee}, {Lenz}, {Littlefair}, {Ma}, {Macleod}, {Mastropietro}, {McCully},
  {Montagnac}, {Morris}, {Mueller}, {Mumford}, {Muna}, {Murphy}, {Nelson},
  {Nguyen}, {Ninan}, {N{\"o}the}, {Ogaz}, {Oh}, {Parejko}, {Parley}, {Pascual},
  {Patil}, {Patil}, {Plunkett}, {Prochaska}, {Rastogi}, {Reddy Janga},
  {Sabater}, {Sakurikar}, {Seifert}, {Sherbert}, {Sherwood-Taylor}, {Shih},
  {Sick}, {Silbiger}, {Singanamalla}, {Singer}, {Sladen}, {Sooley},
  {Sornarajah}, {Streicher}, {Teuben}, {Thomas}, {Tremblay}, {Turner},
  {Terr{\'o}n}, {van Kerkwijk}, {de la Vega}, {Watkins}, {Weaver}, {Whitmore},
  {Woillez}, {Zabalza}, \& {Astropy Contributors}}]{astropyb}
{Astropy Collaboration}, {Price-Whelan}, A.~M., {Sip{\H o}cz}, B.~M., {et~al.}
  2018, \aj, 156, 123

\bibitem[{{Astropy Collaboration} {et~al.}(2013){Astropy Collaboration},
  {Robitaille}, {Tollerud}, {Greenfield}, {Droettboom}, {Bray}, {Aldcroft},
  {Davis}, {Ginsburg}, {Price-Whelan}, {Kerzendorf}, {Conley}, {Crighton},
  {Barbary}, {Muna}, {Ferguson}, {Grollier}, {Parikh}, {Nair}, {Unther},
  {Deil}, {Woillez}, {Conseil}, {Kramer}, {Turner}, {Singer}, {Fox}, {Weaver},
  {Zabalza}, {Edwards}, {Azalee Bostroem}, {Burke}, {Casey}, {Crawford},
  {Dencheva}, {Ely}, {Jenness}, {Labrie}, {Lim}, {Pierfederici}, {Pontzen},
  {Ptak}, {Refsdal}, {Servillat}, \& {Streicher}}]{astropya}
{Astropy Collaboration}, {Robitaille}, T.~P., {Tollerud}, E.~J., {et~al.} 2013,
  \aap, 558, A33

\bibitem[{{Auger} {et~al.}(2010){Auger}, {Treu}, {Gavazzi}, {Bolton},
  {Koopmans}, \& {Marshall}}]{auger}
{Auger}, M.~W., {Treu}, T., {Gavazzi}, R., {et~al.} 2010, \apjl, 721, L163

\bibitem[{{Bacon} {et~al.}(2010){Bacon}, {Accardo}, {Adjali}, {Anwand},
  {Bauer}, {Biswas}, {Blaizot}, {Boudon}, {Brau-Nogue}, {Brinchmann},
  {Caillier}, {Capoani}, {Carollo}, {Contini}, {Couderc}, {Daguis{\'e}},
  {Deiries}, {Delabre}, {Dreizler}, {Dubois}, {Dupieux}, {Dupuy}, {Emsellem},
  {Fechner}, {Fleischmann}, {Fran{\c c}ois}, {Gallou}, {Gharsa}, {Glindemann},
  {Gojak}, {Guiderdoni}, {Hansali}, {Hahn}, {Jarno}, {Kelz}, {Koehler},
  {Kosmalski}, {Laurent}, {Le Floch}, {Lilly}, {Lizon}, {Loupias}, {Manescau},
  {Monstein}, {Nicklas}, {Olaya}, {Pares}, {Pasquini}, {P{\'e}contal-Rousset},
  {Pell{\'o}}, {Petit}, {Popow}, {Reiss}, {Remillieux}, {Renault}, {Roth},
  {Rupprecht}, {Serre}, {Schaye}, {Soucail}, {Steinmetz}, {Streicher}, {Stuik},
  {Valentin}, {Vernet}, {Weilbacher}, {Wisotzki}, \& {Yerle}}]{Bacon10}
{Bacon}, R., {Accardo}, M., {Adjali}, L., {et~al.} 2010, in \procspie, Vol.
  7735, Ground-based and Airborne Instrumentation for Astronomy III, 773508

\bibitem[{{Barber} {et~al.}(2018{\natexlab{a}}){Barber}, {Crain}, \&
  {Schaye}}]{Barber18}
{Barber}, C., {Crain}, R.~A., \& {Schaye}, J. 2018{\natexlab{a}}, \mnras, 479,
  5448

\bibitem[{{Barber} {et~al.}(2018{\natexlab{b}}){Barber}, {Schaye}, \&
  {Crain}}]{Barber18b}
{Barber}, C., {Schaye}, J., \& {Crain}, R.~A. 2018{\natexlab{b}}, ArXiv
  e-prints [\eprint[arXiv]{1807.11310}]

\bibitem[{{Bastian} {et~al.}(2010){Bastian}, {Covey}, \& {Meyer}}]{bastian}
{Bastian}, N., {Covey}, K.~R., \& {Meyer}, M.~R. 2010, \araa, 48, 339

\bibitem[{{Bernardi} {et~al.}(2018){Bernardi}, {Sheth}, {Fischer}, {Meert},
  {Chae}, {Dominguez-Sanchez}, {Huertas-Company}, {Shankar}, \&
  {Vikram}}]{Bernardi18}
{Bernardi}, M., {Sheth}, R.~K., {Fischer}, J.-L., {et~al.} 2018, \mnras, 475,
  757

\bibitem[{{Blakeslee} {et~al.}(2009){Blakeslee}, {Jord{\'a}n}, {Mei},
  {C{\^o}t{\'e}}, {Ferrarese}, {Infante}, {Peng}, {Tonry}, \&
  {West}}]{Blakeslee09}
{Blakeslee}, J.~P., {Jord{\'a}n}, A., {Mei}, S., {et~al.} 2009, \apj, 694, 556

\bibitem[{{Burstein} {et~al.}(1984){Burstein}, {Faber}, {Gaskell}, \&
  {Krumm}}]{Burstein84}
{Burstein}, D., {Faber}, S.~M., {Gaskell}, C.~M., \& {Krumm}, N. 1984, \apj,
  287, 586

\bibitem[{{Cappellari}(2002)}]{Cappellari02}
{Cappellari}, M. 2002, \mnras, 333, 400

\bibitem[{{Cappellari}(2017)}]{Cappellari17}
{Cappellari}, M. 2017, \mnras, 466, 798

\bibitem[{{Cappellari} \& {Copin}(2003)}]{voronoi}
{Cappellari}, M. \& {Copin}, Y. 2003, \mnras, 342, 345

\bibitem[{{Cappellari} \& {Emsellem}(2004)}]{ppxf}
{Cappellari}, M. \& {Emsellem}, E. 2004, \pasp, 116, 138

\bibitem[{{Cappellari} {et~al.}(2012){Cappellari}, {McDermid}, {Alatalo},
  {Blitz}, {Bois}, {Bournaud}, {Bureau}, {Crocker}, {Davies}, {Davis}, {de
  Zeeuw}, {Duc}, {Emsellem}, {Khochfar}, {Krajnovi{\'c}}, {Kuntschner},
  {Lablanche}, {Morganti}, {Naab}, {Oosterloo}, {Sarzi}, {Scott}, {Serra},
  {Weijmans}, \& {Young}}]{cappellari}
{Cappellari}, M., {McDermid}, R.~M., {Alatalo}, K., {et~al.} 2012, \nat, 484,
  485

\bibitem[{{Cenarro} {et~al.}(2001){Cenarro}, {Cardiel}, {Gorgas}, {Peletier},
  {Vazdekis}, \& {Prada}}]{cat}
{Cenarro}, A.~J., {Cardiel}, N., {Gorgas}, J., {et~al.} 2001, \mnras, 326, 959

\bibitem[{{Chabrier}(2003)}]{Chabrier}
{Chabrier}, G. 2003, \pasp, 115, 763

\bibitem[{{Chabrier} {et~al.}(2014){Chabrier}, {Hennebelle}, \&
  {Charlot}}]{Chabrier14}
{Chabrier}, G., {Hennebelle}, P., \& {Charlot}, S. 2014, \apj, 796, 75

\bibitem[{{Cid Fernandes} {et~al.}(2005){Cid Fernandes}, {Mateus}, {Sodr{\'e}},
  {Stasi{\'n}ska}, \& {Gomes}}]{CF05}
{Cid Fernandes}, R., {Mateus}, A., {Sodr{\'e}}, L., {Stasi{\'n}ska}, G., \&
  {Gomes}, J.~M. 2005, \mnras, 358, 363

\bibitem[{{Clauwens} {et~al.}(2016){Clauwens}, {Schaye}, \&
  {Franx}}]{Clauwens16}
{Clauwens}, B., {Schaye}, J., \& {Franx}, M. 2016, \mnras, 462, 2832

\bibitem[{{Conroy} {et~al.}(2014){Conroy}, {Graves}, \& {van
  Dokkum}}]{Conroy14}
{Conroy}, C., {Graves}, G.~J., \& {van Dokkum}, P.~G. 2014, \apj, 780, 33

\bibitem[{{Conroy} {et~al.}(2009){Conroy}, {Gunn}, \& {White}}]{Conroy09}
{Conroy}, C., {Gunn}, J.~E., \& {White}, M. 2009, \apj, 699, 486

\bibitem[{{Conroy} \& {van Dokkum}(2012{\natexlab{a}})}]{conroy}
{Conroy}, C. \& {van Dokkum}, P. 2012{\natexlab{a}}, \apj, 747, 69

\bibitem[{{Conroy} \& {van Dokkum}(2012{\natexlab{b}})}]{conroy12}
{Conroy}, C. \& {van Dokkum}, P.~G. 2012{\natexlab{b}}, \apj, 760, 71

\bibitem[{{Conroy} {et~al.}(2017){Conroy}, {van Dokkum}, \&
  {Villaume}}]{conroy17}
{Conroy}, C., {van Dokkum}, P.~G., \& {Villaume}, A. 2017, \apj, 837, 166

\bibitem[{{Conroy} {et~al.}(2018){Conroy}, {Villaume}, {van Dokkum}, \&
  {Lind}}]{Conroy18}
{Conroy}, C., {Villaume}, A., {van Dokkum}, P.~G., \& {Lind}, K. 2018, \apj,
  854, 139

\bibitem[{{Corsini} {et~al.}(2017){Corsini}, {Wegner}, {Thomas}, {Saglia}, \&
  {Bender}}]{Corsini17}
{Corsini}, E.~M., {Wegner}, G.~A., {Thomas}, J., {Saglia}, R.~P., \& {Bender},
  R. 2017, \mnras, 466, 974

\bibitem[{{Courteau} {et~al.}(2014){Courteau}, {Cappellari}, {de Jong},
  {Dutton}, {Emsellem}, {Hoekstra}, {Koopmans}, {Mamon}, {Maraston}, {Treu}, \&
  {Widrow}}]{Courteau14}
{Courteau}, S., {Cappellari}, M., {de Jong}, R.~S., {et~al.} 2014, Reviews of
  Modern Physics, 86, 47

\bibitem[{{Davis} \& {McDermid}(2017)}]{Davis17}
{Davis}, T.~A. \& {McDermid}, R.~M. 2017, \mnras, 464, 453

\bibitem[{{de La Rosa} {et~al.}(2011){de La Rosa}, {La Barbera}, {Ferreras}, \&
  {de Carvalho}}]{dlr11}
{de La Rosa}, I.~G., {La Barbera}, F., {Ferreras}, I., \& {de Carvalho}, R.~R.
  2011, \mnras, 418, L74

\bibitem[{{De Masi} {et~al.}(2018){De Masi}, {Vincenzo}, {Matteucci}, {Rosani},
  {Barbera}, {Pasquali}, \& {Spitoni}}]{DM18}
{De Masi}, C., {Vincenzo}, F., {Matteucci}, F., {et~al.} 2018, ArXiv e-prints
  [\eprint[arXiv]{1805.06841}]

\bibitem[{{Drinkwater} {et~al.}(2001){Drinkwater}, {Gregg}, \&
  {Colless}}]{Drinkwater01}
{Drinkwater}, M.~J., {Gregg}, M.~D., \& {Colless}, M. 2001, \apjl, 548, L139

\bibitem[{{Dutton} {et~al.}(2012){Dutton}, {Mendel}, \& {Simard}}]{Dutton12}
{Dutton}, A.~A., {Mendel}, J.~T., \& {Simard}, L. 2012, \mnras, 422, L33

\bibitem[{{Emsellem} {et~al.}(1994){Emsellem}, {Monnet}, \&
  {Bacon}}]{Emsellem94}
{Emsellem}, E., {Monnet}, G., \& {Bacon}, R. 1994, \aap, 285, 723

\bibitem[{{Erwin}(2015)}]{Erwin15}
{Erwin}, P. 2015, \apj, 799, 226

\bibitem[{{Falc{\'o}n-Barroso} {et~al.}(2011){Falc{\'o}n-Barroso},
  {S{\'a}nchez-Bl{\'a}zquez}, {Vazdekis}, {Ricciardelli}, {Cardiel}, {Cenarro},
  {Gorgas}, \& {Peletier}}]{Jesus11}
{Falc{\'o}n-Barroso}, J., {S{\'a}nchez-Bl{\'a}zquez}, P., {Vazdekis}, A.,
  {et~al.} 2011, \aap, 532, A95

\bibitem[{{Ferguson}(1989)}]{Ferguson89}
{Ferguson}, H.~C. 1989, \aj, 98, 367

\bibitem[{{Fern{\'a}ndez-Alvar} {et~al.}(2018){Fern{\'a}ndez-Alvar}, {Tissera},
  {Carigi}, {Schuster}, {Beers}, \& {Belokurov}}]{Emma18}
{Fern{\'a}ndez-Alvar}, E., {Tissera}, P.~B., {Carigi}, L., {et~al.} 2018, ArXiv
  e-prints [\eprint[arXiv]{1809.02368}]

\bibitem[{{Ferr{\'e}-Mateu} {et~al.}(2013){Ferr{\'e}-Mateu}, {Vazdekis}, \& {de
  la Rosa}}]{FM13}
{Ferr{\'e}-Mateu}, A., {Vazdekis}, A., \& {de la Rosa}, I.~G. 2013, \mnras,
  431, 440

\bibitem[{{Ferreras} {et~al.}(2013){Ferreras}, {La Barbera}, {de la Rosa},
  {Vazdekis}, {de Carvalho}, {Falc{\'o}n-Barroso}, \&
  {Ricciardelli}}]{ferreras}
{Ferreras}, I., {La Barbera}, F., {de la Rosa}, I.~G., {et~al.} 2013, \mnras,
  429, L15

\bibitem[{{Ferreras} {et~al.}(2015){Ferreras}, {Weidner}, {Vazdekis}, \& {La
  Barbera}}]{Ferreras15}
{Ferreras}, I., {Weidner}, C., {Vazdekis}, A., \& {La Barbera}, F. 2015,
  \mnras, 448, L82

\bibitem[{{Fontanot} {et~al.}(2018){Fontanot}, {La Barbera}, {De Lucia},
  {Pasquali}, \& {Vazdekis}}]{Fontanot18}
{Fontanot}, F., {La Barbera}, F., {De Lucia}, G., {Pasquali}, A., \&
  {Vazdekis}, A. 2018, \mnras, 479, 5678

\bibitem[{{Foreman-Mackey} {et~al.}(2013){Foreman-Mackey}, {Hogg}, {Lang}, \&
  {Goodman}}]{emcee}
{Foreman-Mackey}, D., {Hogg}, D.~W., {Lang}, D., \& {Goodman}, J. 2013, \pasp,
  125, 306

\bibitem[{{Freudling} {et~al.}(2013){Freudling}, {Romaniello}, {Bramich},
  {Ballester}, {Forchi}, {Garc{\'{\i}}a-Dabl{\'o}}, {Moehler}, \&
  {Neeser}}]{Freudling13}
{Freudling}, W., {Romaniello}, M., {Bramich}, D.~M., {et~al.} 2013, \aap, 559,
  A96

\bibitem[{{Garrison-Kimmel} {et~al.}(2018){Garrison-Kimmel}, {Hopkins},
  {Wetzel}, {El-Badry}, {Sanderson}, {Bullock}, {Ma}, {van de Voort}, {Hafen},
  {Faucher-Gigu{\`e}re}, {Hayward}, {Quataert}, {Kere{\v{s}}}, \&
  {Boylan-Kolchin}}]{GK18}
{Garrison-Kimmel}, S., {Hopkins}, P.~F., {Wetzel}, A., {et~al.} 2018, \mnras,
  481, 4133

\bibitem[{{Grand} {et~al.}(2016){Grand}, {Springel}, {G{\'o}mez}, {Marinacci},
  {Pakmor}, {Campbell}, \& {Jenkins}}]{Grand16}
{Grand}, R.~J.~J., {Springel}, V., {G{\'o}mez}, F.~A., {et~al.} 2016, \mnras,
  459, 199

\bibitem[{{Gutcke} \& {Springel}(2018)}]{Thales18}
{Gutcke}, T.~A. \& {Springel}, V. 2018, \mnras [\eprint[arXiv]{1710.04222}]

\bibitem[{{Hayes} {et~al.}(2018){Hayes}, {Majewski}, {Shetrone},
  {Fern{\'a}ndez-Alvar}, {Allende Prieto}, {Schuster}, {Carigi}, {Cunha},
  {Smith}, {Sobeck}, {Almeida}, {Beers}, {Carrera}, {Fern{\'a}ndez-Trincado},
  {Garc{\'{\i}}a-Hern{\'a}ndez}, {Geisler}, {Lane}, {Lucatello}, {Matthews},
  {Minniti}, {Nitschelm}, {Tang}, {Tissera}, \& {Zamora}}]{Hayes18}
{Hayes}, C.~R., {Majewski}, S.~R., {Shetrone}, M., {et~al.} 2018, \apj, 852, 49

\bibitem[{{Iodice} {et~al.}(2018){Iodice}, {Spavone}, {Capaccioli}, {Peletier},
  {van de Ven}, {Napolitano}, {Hilker}, {Mieske}, {Smith}, {Pasquali},
  {Limatola}, {Grado}, {Venhola}, {Cantiello}, {Paolillo}, {Falcon-Barroso},
  {D'Abrusco}, \& {Schipani}}]{Enrica18}
{Iodice}, E., {Spavone}, M., {Capaccioli}, M., {et~al.} 2018, arXiv e-prints,
  arXiv:1812.01050

\bibitem[{{Jerabkova} {et~al.}(2018){Jerabkova}, {Zonoozi}, {Kroupa},
  {Beccari}, {Yan}, {Vazdekis}, \& {Zhang}}]{Jerabkova18}
{Jerabkova}, T., {Zonoozi}, A.~H., {Kroupa}, P., {et~al.} 2018, ArXiv e-prints
  [\eprint[arXiv]{1809.04603}]

\bibitem[{{Johansson} {et~al.}(2012){Johansson}, {Thomas}, \&
  {Maraston}}]{johansson12}
{Johansson}, J., {Thomas}, D., \& {Maraston}, C. 2012, \mnras, 421, 1908

\bibitem[{{Jorgensen}(1994)}]{Jorgensen94}
{Jorgensen}, U.~G. 1994, \aap, 284, 179

\bibitem[{{Kennicutt}(1998)}]{Kennicutt98}
{Kennicutt}, Jr., R.~C. 1998, \araa, 36, 189

\bibitem[{{Kroupa}(2001)}]{mw}
{Kroupa}, P. 2001, \mnras, 322, 231

\bibitem[{{Kroupa}(2002)}]{Kroupa}
{Kroupa}, P. 2002, Science, 295, 82

\bibitem[{{Kuntschner} {et~al.}(2010){Kuntschner}, {Emsellem}, {Bacon},
  {Cappellari}, {Davies}, {de Zeeuw}, {Falc{\'o}n-Barroso}, {Krajnovi{\'c}},
  {McDermid}, {Peletier}, {Sarzi}, {Shapiro}, {van den Bosch}, \& {van de
  Ven}}]{Kuntschner10}
{Kuntschner}, H., {Emsellem}, E., {Bacon}, R., {et~al.} 2010, \mnras, 408, 97

\bibitem[{{La Barbera} {et~al.}(2015){La Barbera}, {Ferreras}, \&
  {Vazdekis}}]{LB15}
{La Barbera}, F., {Ferreras}, I., \& {Vazdekis}, A. 2015, \mnras, 449, L137

\bibitem[{{La Barbera} {et~al.}(2013){La Barbera}, {Ferreras}, {Vazdekis}, {de
  la Rosa}, {de Carvalho}, {Trevisan}, {Falc{\'o}n-Barroso}, \&
  {Ricciardelli}}]{labarbera}
{La Barbera}, F., {Ferreras}, I., {Vazdekis}, A., {et~al.} 2013, \mnras, 433,
  3017

\bibitem[{{La Barbera} {et~al.}(2014){La Barbera}, {Pasquali}, {Ferreras},
  {Gallazzi}, {de Carvalho}, \& {de la Rosa}}]{LB14}
{La Barbera}, F., {Pasquali}, A., {Ferreras}, I., {et~al.} 2014, \mnras, 445,
  1977

\bibitem[{{La Barbera} {et~al.}(2016){La Barbera}, {Vazdekis}, {Ferreras},
  {Pasquali}, {Cappellari}, {Mart{\'{\i}}n-Navarro}, {Sch{\"o}nebeck}, \&
  {Falc{\'o}n-Barroso}}]{LB16}
{La Barbera}, F., {Vazdekis}, A., {Ferreras}, I., {et~al.} 2016, \mnras, 457,
  1468

\bibitem[{{L{\"a}sker} {et~al.}(2013){L{\"a}sker}, {van den Bosch}, {van de
  Ven}, {Ferreras}, {La Barbera}, {Vazdekis}, \&
  {Falc{\'o}n-Barroso}}]{Lasker13}
{L{\"a}sker}, R., {van den Bosch}, R.~C.~E., {van de Ven}, G., {et~al.} 2013,
  \mnras, 434, L31

\bibitem[{{Lyubenova} {et~al.}(2016){Lyubenova}, {Mart{\'{\i}}n-Navarro}, {van
  de Ven}, {Falc{\'o}n-Barroso}, {Galbany}, {Gallazzi}, {Garc{\'{\i}}a-Benito},
  {Gonz{\'a}lez Delgado}, {Husemann}, {La Barbera}, {Marino}, {Mast},
  {Mendez-Abreu}, {Peletier}, {S{\'a}nchez-Bl{\'a}zquez}, {S{\'a}nchez},
  {Trager}, {van den Bosch}, {Vazdekis}, {Walcher}, {Zhu}, {Zibetti},
  {Ziegler}, {Bland-Hawthorn}, \& {CALIFA Collaboration}}]{Lyubenova16}
{Lyubenova}, M., {Mart{\'{\i}}n-Navarro}, I., {van de Ven}, G., {et~al.} 2016,
  \mnras, 463, 3220

\bibitem[{{Madau} \& {Dickinson}(2014)}]{Madau14}
{Madau}, P. \& {Dickinson}, M. 2014, \araa, 52, 415

\bibitem[{{Mart{\'{\i}}n-Navarro}(2016)}]{MN16}
{Mart{\'{\i}}n-Navarro}, I. 2016, \mnras, 456, L104

\bibitem[{{Mart{\'{\i}}n-Navarro}
  {et~al.}(2015{\natexlab{a}}){Mart{\'{\i}}n-Navarro}, {La Barbera},
  {Vazdekis}, {Falc{\'o}n-Barroso}, \& {Ferreras}}]{MN15a}
{Mart{\'{\i}}n-Navarro}, I., {La Barbera}, F., {Vazdekis}, A.,
  {Falc{\'o}n-Barroso}, J., \& {Ferreras}, I. 2015{\natexlab{a}}, \mnras, 447,
  1033

\bibitem[{{Mart{\'{\i}}n-Navarro}
  {et~al.}(2015{\natexlab{b}}){Mart{\'{\i}}n-Navarro}, {La Barbera},
  {Vazdekis}, {Ferr{\'e}-Mateu}, {Trujillo}, \& {Beasley}}]{MN15b}
{Mart{\'{\i}}n-Navarro}, I., {La Barbera}, F., {Vazdekis}, A., {et~al.}
  2015{\natexlab{b}}, \mnras, 451, 1081

\bibitem[{{Mart{\'{\i}}n-Navarro}
  {et~al.}(2015{\natexlab{c}}){Mart{\'{\i}}n-Navarro},
  {P{\'e}rez-Gonz{\'a}lez}, {Trujillo}, {Esquej}, {Vazdekis}, {Dom{\'{\i}}nguez
  S{\'a}nchez}, {Barro}, {Bruzual}, {Charlot}, {Cava}, {Ferreras}, {Espino},
  {La Barbera}, {Koekemoer}, \& {Cenarro}}]{MN15d}
{Mart{\'{\i}}n-Navarro}, I., {P{\'e}rez-Gonz{\'a}lez}, P.~G., {Trujillo}, I.,
  {et~al.} 2015{\natexlab{c}}, \apjl, 798, L4

\bibitem[{{Mart{\'{\i}}n-Navarro} {et~al.}(2018){Mart{\'{\i}}n-Navarro},
  {Vazdekis}, {Falc{\'o}n-Barroso}, {La Barbera}, {Y{\i}ld{\i}r{\i}m}, \& {van
  de Ven}}]{MN18}
{Mart{\'{\i}}n-Navarro}, I., {Vazdekis}, A., {Falc{\'o}n-Barroso}, J., {et~al.}
  2018, \mnras, 475, 3700

\bibitem[{{Mart{\'{\i}}n-Navarro}
  {et~al.}(2015{\natexlab{d}}){Mart{\'{\i}}n-Navarro}, {Vazdekis}, {La
  Barbera}, {Falc{\'o}n-Barroso}, {Lyubenova}, {van de Ven}, {Ferreras},
  {S{\'a}nchez}, {Trager}, {Garc{\'{\i}}a-Benito}, {Mast}, {Mendoza},
  {S{\'a}nchez-Bl{\'a}zquez}, {Gonz{\'a}lez Delgado}, {Walcher}, \& {The CALIFA
  Team}}]{MN15c}
{Mart{\'{\i}}n-Navarro}, I., {Vazdekis}, A., {La Barbera}, F., {et~al.}
  2015{\natexlab{d}}, \apjl, 806, L31

\bibitem[{{McConnell} {et~al.}(2016){McConnell}, {Lu}, \& {Mann}}]{McConnell16}
{McConnell}, N.~J., {Lu}, J.~R., \& {Mann}, A.~W. 2016, \apj, 821, 39

\bibitem[{{McDermid} {et~al.}(2015){McDermid}, {Alatalo}, {Blitz}, {Bournaud},
  {Bureau}, {Cappellari}, {Crocker}, {Davies}, {Davis}, {de Zeeuw}, {Duc},
  {Emsellem}, {Khochfar}, {Krajnovi{\'c}}, {Kuntschner}, {Morganti}, {Naab},
  {Oosterloo}, {Sarzi}, {Scott}, {Serra}, {Weijmans}, \& {Young}}]{McDermid15}
{McDermid}, R.~M., {Alatalo}, K., {Blitz}, L., {et~al.} 2015, \mnras, 448, 3484

\bibitem[{{McGee} {et~al.}(2014){McGee}, {Goto}, \& {Balogh}}]{McGee14}
{McGee}, S.~L., {Goto}, R., \& {Balogh}, M.~L. 2014, \mnras, 438, 3188

\bibitem[{{Miller} \& {Scalo}(1979)}]{Miller79}
{Miller}, G.~E. \& {Scalo}, J.~M. 1979, \apjs, 41, 513

\bibitem[{{Milone} {et~al.}(2011){Milone}, {Sansom}, \&
  {S{\'a}nchez-Bl{\'a}zquez}}]{Milone11}
{Milone}, A.~D.~C., {Sansom}, A.~E., \& {S{\'a}nchez-Bl{\'a}zquez}, P. 2011,
  \mnras, 414, 1227

\bibitem[{{Mitchell} {et~al.}(2013){Mitchell}, {Lacey}, {Baugh}, \&
  {Cole}}]{Mitchell13}
{Mitchell}, P.~D., {Lacey}, C.~G., {Baugh}, C.~M., \& {Cole}, S. 2013, \mnras,
  435, 87

\bibitem[{{Molaeinezhad} {et~al.}(2017){Molaeinezhad}, {Falc{\'o}n-Barroso},
  {Mart{\'{\i}}nez-Valpuesta}, {Khosroshahi}, {Vazdekis}, {La Barbera},
  {Peletier}, \& {Balcells}}]{Molaeinezhad17}
{Molaeinezhad}, A., {Falc{\'o}n-Barroso}, J., {Mart{\'{\i}}nez-Valpuesta}, I.,
  {et~al.} 2017, \mnras, 467, 353

\bibitem[{{Newman} {et~al.}(2017){Newman}, {Smith}, {Conroy}, {Villaume}, \&
  {van Dokkum}}]{Newman17}
{Newman}, A.~B., {Smith}, R.~J., {Conroy}, C., {Villaume}, A., \& {van Dokkum},
  P. 2017, \apj, 845, 157

\bibitem[{{Ocvirk} {et~al.}(2006){Ocvirk}, {Pichon}, {Lan{\c c}on}, \&
  {Thi{\'e}baut}}]{Ocvirk06}
{Ocvirk}, P., {Pichon}, C., {Lan{\c c}on}, A., \& {Thi{\'e}baut}, E. 2006,
  \mnras, 365, 74

\bibitem[{{Oldham} \& {Auger}(2018)}]{Oldham}
{Oldham}, L. \& {Auger}, M. 2018, \mnras, 474, 4169

\bibitem[{{Parikh} {et~al.}(2018){Parikh}, {Thomas}, {Maraston}, {Westfall},
  {Goddard}, {Lian}, {Meneses-Goytia}, {Jones}, {Vaughan}, {Andrews},
  {Bershady}, {Bizyaev}, {Brinkmann}, {Brownstein}, {Bundy}, {Drory},
  {Emsellem}, {Law}, {Newman}, {Roman-Lopes}, {Wake}, {Yan}, \&
  {Zheng}}]{Parikh}
{Parikh}, T., {Thomas}, D., {Maraston}, C., {et~al.} 2018, \mnras, 477, 3954

\bibitem[{{Philcox} {et~al.}(2018){Philcox}, {Rybizki}, \&
  {Gutcke}}]{Philcox18}
{Philcox}, O., {Rybizki}, J., \& {Gutcke}, T.~A. 2018, \apj, 861, 40

\bibitem[{{Pietrinferni} {et~al.}(2004){Pietrinferni}, {Cassisi}, {Salaris}, \&
  {Castelli}}]{basti1}
{Pietrinferni}, A., {Cassisi}, S., {Salaris}, M., \& {Castelli}, F. 2004, \apj,
  612, 168

\bibitem[{{Pietrinferni} {et~al.}(2006){Pietrinferni}, {Cassisi}, {Salaris}, \&
  {Castelli}}]{basti2}
{Pietrinferni}, A., {Cassisi}, S., {Salaris}, M., \& {Castelli}, F. 2006, \apj,
  642, 797

\bibitem[{{Pinna} {et~al.}(2019){Pinna}, {Falc{\'o}n-Barroso}, {Martig},
  {Sarzi}, {Coccato}, {Iodice}, {Corsini}, {de Zeeuw}, {Gadotti}, {Leaman},
  {Lyubenova}, {McDermid}, {Minchev}, {Morelli}, {van de Ven}, \&
  {Viaene}}]{Pinna19}
{Pinna}, F., {Falc{\'o}n-Barroso}, J., {Martig}, M., {et~al.} 2019, arXiv
  e-prints [\eprint[arXiv]{1901.04310}]

\bibitem[{{Salpeter}(1955)}]{Salp:55}
{Salpeter}, E.~E. 1955, \apj, 121, 161

\bibitem[{{S{\'a}nchez} {et~al.}(2012){S{\'a}nchez}, {Kennicutt}, {Gil de Paz},
  {van de Ven}, {V{\'{\i}}lchez}, {Wisotzki}, {Walcher}, {Mast}, {Aguerri},
  {Albiol-P{\'e}rez}, {Alonso-Herrero}, {Alves}, {Bakos}, {Bart{\'a}kov{\'a}},
  {Bland-Hawthorn}, {Boselli}, {Bomans}, {Castillo-Morales}, {Cortijo-Ferrero},
  {de Lorenzo-C{\'a}ceres}, {Del Olmo}, {Dettmar}, {D{\'{\i}}az}, {Ellis},
  {Falc{\'o}n-Barroso}, {Flores}, {Gallazzi}, {Garc{\'{\i}}a-Lorenzo},
  {Gonz{\'a}lez Delgado}, {Gruel}, {Haines}, {Hao}, {Husemann},
  {Igl{\'e}sias-P{\'a}ramo}, {Jahnke}, {Johnson}, {Jungwiert}, {Kalinova},
  {Kehrig}, {Kupko}, {L{\'o}pez-S{\'a}nchez}, {Lyubenova}, {Marino},
  {M{\'a}rmol-Queralt{\'o}}, {M{\'a}rquez}, {Masegosa}, {Meidt},
  {Mendez-Abreu}, {Monreal-Ibero}, {Montijo}, {Mour{\~a}o}, {Palacios-Navarro},
  {Papaderos}, {Pasquali}, {Peletier}, {P{\'e}rez}, {P{\'e}rez}, {Quirrenbach},
  {Rela{\~n}o}, {Rosales-Ortega}, {Roth}, {Ruiz-Lara},
  {S{\'a}nchez-Bl{\'a}zquez}, {Sengupta}, {Singh}, {Stanishev}, {Trager},
  {Vazdekis}, {Viironen}, {Wild}, {Zibetti}, \& {Ziegler}}]{califa}
{S{\'a}nchez}, S.~F., {Kennicutt}, R.~C., {Gil de Paz}, A., {et~al.} 2012,
  \aap, 538, A8

\bibitem[{{S{\'a}nchez-Bl{\'a}zquez} {et~al.}(2011){S{\'a}nchez-Bl{\'a}zquez},
  {Ocvirk}, {Gibson}, {P{\'e}rez}, \& {Peletier}}]{Pat11}
{S{\'a}nchez-Bl{\'a}zquez}, P., {Ocvirk}, P., {Gibson}, B.~K., {P{\'e}rez}, I.,
  \& {Peletier}, R.~F. 2011, \mnras, 415, 709

\bibitem[{{S{\'a}nchez-Bl{\'a}zquez} {et~al.}(2006){S{\'a}nchez-Bl{\'a}zquez},
  {Peletier}, {Jim{\'e}nez-Vicente}, {Cardiel}, {Cenarro},
  {Falc{\'o}n-Barroso}, {Gorgas}, {Selam}, \& {Vazdekis}}]{Pat06}
{S{\'a}nchez-Bl{\'a}zquez}, P., {Peletier}, R.~F., {Jim{\'e}nez-Vicente}, J.,
  {et~al.} 2006, \mnras, 371, 703

\bibitem[{{Sarzi} {et~al.}(2018{\natexlab{a}}){Sarzi}, {Iodice}, {Coccato},
  {Corsini}, {de Zeeuw}, {Falc{\'o}n-Barroso}, {Gadotti}, {Lyubenova},
  {McDermid}, {van de Ven}, {Fahrion}, {Pizzella}, \& {Zhu}}]{f3d}
{Sarzi}, M., {Iodice}, E., {Coccato}, L., {et~al.} 2018{\natexlab{a}}, \aap,
  616, A121

\bibitem[{{Sarzi} {et~al.}(2018{\natexlab{b}}){Sarzi}, {Spiniello}, {La
  Barbera}, {Krajnovi{\'c}}, \& {van den Bosch}}]{Sarzi18}
{Sarzi}, M., {Spiniello}, C., {La Barbera}, F., {Krajnovi{\'c}}, D., \& {van
  den Bosch}, R. 2018{\natexlab{b}}, \mnras, 478, 4084

\bibitem[{{Schiavon}(2007)}]{Schiavon07}
{Schiavon}, R.~P. 2007, \apjs, 171, 146

\bibitem[{{Serven} {et~al.}(2005){Serven}, {Worthey}, \& {Briley}}]{serven}
{Serven}, J., {Worthey}, G., \& {Briley}, M.~M. 2005, \apj, 627, 754

\bibitem[{{Smith}(2014)}]{smith}
{Smith}, R.~J. 2014, \mnras, 443, L69

\bibitem[{{Smith} \& {Lucey}(2013)}]{Smith13}
{Smith}, R.~J. \& {Lucey}, J.~R. 2013, \mnras, 434, 1964

\bibitem[{{Smith} {et~al.}(2012){Smith}, {Lucey}, \& {Carter}}]{Smith12}
{Smith}, R.~J., {Lucey}, J.~R., \& {Carter}, D. 2012, \mnras, 426, 2994

\bibitem[{{Soto} {et~al.}(2016){Soto}, {Lilly}, {Bacon}, {Richard}, \&
  {Conseil}}]{Soto16}
{Soto}, K.~T., {Lilly}, S.~J., {Bacon}, R., {Richard}, J., \& {Conseil}, S.
  2016, \mnras, 458, 3210

\bibitem[{{Spiniello} {et~al.}(2014){Spiniello}, {Trager}, {Koopmans}, \&
  {Conroy}}]{Spiniello2013}
{Spiniello}, C., {Trager}, S., {Koopmans}, L.~V.~E., \& {Conroy}, C. 2014,
  \mnras, 438, 1483

\bibitem[{{Spiniello} {et~al.}(2015){Spiniello}, {Trager}, \&
  {Koopmans}}]{Spiniello15}
{Spiniello}, C., {Trager}, S.~C., \& {Koopmans}, L.~V.~E. 2015, \apj, 803, 87

\bibitem[{{Spiniello} {et~al.}(2012){Spiniello}, {Trager}, {Koopmans}, \&
  {Chen}}]{spiniello12}
{Spiniello}, C., {Trager}, S.~C., {Koopmans}, L.~V.~E., \& {Chen}, Y.~P. 2012,
  \apjl, 753, L32

\bibitem[{{Sybilska} {et~al.}(2017){Sybilska}, {Lisker}, {Kuntschner},
  {Vazdekis}, {van de Ven}, {Peletier}, {Falc{\'o}n-Barroso}, {Vijayaraghavan},
  \& {Janz}}]{Aga17}
{Sybilska}, A., {Lisker}, T., {Kuntschner}, H., {et~al.} 2017, \mnras, 470, 815

\bibitem[{{Tang} \& {Worthey}(2017)}]{Tang17}
{Tang}, B. \& {Worthey}, G. 2017, \mnras, 467, 674

\bibitem[{{Thomas} {et~al.}(1999){Thomas}, {Greggio}, \& {Bender}}]{Thomas99}
{Thomas}, D., {Greggio}, L., \& {Bender}, R. 1999, \mnras, 302, 537

\bibitem[{{Thomas} {et~al.}(2003){Thomas}, {Maraston}, \& {Bender}}]{TMB:03}
{Thomas}, D., {Maraston}, C., \& {Bender}, R. 2003, \mnras, 339, 897

\bibitem[{{Thomas} {et~al.}(2005){Thomas}, {Maraston}, {Bender}, \& {Mendes de
  Oliveira}}]{Thomas05}
{Thomas}, D., {Maraston}, C., {Bender}, R., \& {Mendes de Oliveira}, C. 2005,
  \apj, 621, 673

\bibitem[{{Thomas} {et~al.}(2010){Thomas}, {Maraston}, {Schawinski}, {Sarzi},
  \& {Silk}}]{Thomas10}
{Thomas}, D., {Maraston}, C., {Schawinski}, K., {Sarzi}, M., \& {Silk}, J.
  2010, \mnras, 404, 1775

\bibitem[{{Thomas} {et~al.}(2011){Thomas}, {Saglia}, {Bender}, {Thomas},
  {Gebhardt}, {Magorrian}, {Corsini}, {Wegner}, \& {Seitz}}]{thomas11}
{Thomas}, J., {Saglia}, R.~P., {Bender}, R., {et~al.} 2011, \mnras, 415, 545

\bibitem[{{Tortora} {et~al.}(2014){Tortora}, {La Barbera}, {Napolitano},
  {Romanowsky}, {Ferreras}, \& {de Carvalho}}]{Tortora14}
{Tortora}, C., {La Barbera}, F., {Napolitano}, N.~R., {et~al.} 2014, \mnras,
  445, 115

\bibitem[{{Tortora} {et~al.}(2013){Tortora}, {Romanowsky}, \&
  {Napolitano}}]{Tortora13b}
{Tortora}, C., {Romanowsky}, A.~J., \& {Napolitano}, N.~R. 2013, \apj, 765, 8

\bibitem[{{Trager} {et~al.}(1998){Trager}, {Worthey}, {Faber}, {Burstein}, \&
  {Gonzalez}}]{trager}
{Trager}, S.~C., {Worthey}, G., {Faber}, S.~M., {Burstein}, D., \& {Gonzalez},
  J.~J. 1998, \apjs, 116, 1

\bibitem[{{Treu} {et~al.}(2010){Treu}, {Auger}, {Koopmans}, {Gavazzi},
  {Marshall}, \& {Bolton}}]{Treu}
{Treu}, T., {Auger}, M.~W., {Koopmans}, L.~V.~E., {et~al.} 2010, \apj, 709,
  1195

\bibitem[{{van den Bosch} {et~al.}(2008){van den Bosch}, {van de Ven},
  {Verolme}, {Cappellari}, \& {de Zeeuw}}]{remco08}
{van den Bosch}, R.~C.~E., {van de Ven}, G., {Verolme}, E.~K., {Cappellari},
  M., \& {de Zeeuw}, P.~T. 2008, \mnras, 385, 647

\bibitem[{{van Dokkum} {et~al.}(2017){van Dokkum}, {Conroy}, {Villaume},
  {Brodie}, \& {Romanowsky}}]{vdk17}
{van Dokkum}, P., {Conroy}, C., {Villaume}, A., {Brodie}, J., \& {Romanowsky},
  A.~J. 2017, \apj, 841, 68

\bibitem[{{van Dokkum} \& {Conroy}(2010)}]{vandokkum}
{van Dokkum}, P.~G. \& {Conroy}, C. 2010, \nat, 468, 940

\bibitem[{{Vargas} {et~al.}(2014){Vargas}, {Gilbert}, {Geha}, {Tollerud},
  {Kirby}, \& {Guhathakurta}}]{Vargas14}
{Vargas}, L.~C., {Gilbert}, K.~M., {Geha}, M., {et~al.} 2014, \apjl, 797, L2

\bibitem[{{Vaughan} {et~al.}(2018{\natexlab{a}}){Vaughan}, {Davies},
  {Zieleniewski}, \& {Houghton}}]{Vaughan18b}
{Vaughan}, S.~P., {Davies}, R.~L., {Zieleniewski}, S., \& {Houghton}, R.~C.~W.
  2018{\natexlab{a}}, \mnras, 475, 1073

\bibitem[{{Vaughan} {et~al.}(2018{\natexlab{b}}){Vaughan}, {Davies},
  {Zieleniewski}, \& {Houghton}}]{Vaughan18a}
{Vaughan}, S.~P., {Davies}, R.~L., {Zieleniewski}, S., \& {Houghton}, R.~C.~W.
  2018{\natexlab{b}}, \mnras, 479, 2443

\bibitem[{{Vazdekis} {et~al.}(1996){Vazdekis}, {Casuso}, {Peletier}, \&
  {Beckman}}]{vazdekis96}
{Vazdekis}, A., {Casuso}, E., {Peletier}, R.~F., \& {Beckman}, J.~E. 1996,
  \apjs, 106, 307

\bibitem[{{Vazdekis} {et~al.}(2015){Vazdekis}, {Coelho}, {Cassisi},
  {Ricciardelli}, {Falc{\'o}n-Barroso}, {S{\'a}nchez-Bl{\'a}zquez}, {La
  Barbera}, {Beasley}, \& {Pietrinferni}}]{Vazdekis15}
{Vazdekis}, A., {Coelho}, P., {Cassisi}, S., {et~al.} 2015, \mnras, 449, 1177

\bibitem[{{Vazdekis} {et~al.}(2016){Vazdekis}, {Koleva}, {Ricciardelli},
  {R{\"o}ck}, \& {Falc{\'o}n-Barroso}}]{Vazdekis16}
{Vazdekis}, A., {Koleva}, M., {Ricciardelli}, E., {R{\"o}ck}, B., \&
  {Falc{\'o}n-Barroso}, J. 2016, \mnras, 463, 3409

\bibitem[{{Vazdekis} {et~al.}(2010){Vazdekis}, {S{\'a}nchez-Bl{\'a}zquez},
  {Falc{\'o}n-Barroso}, {Cenarro}, {Beasley}, {Cardiel}, {Gorgas}, \&
  {Peletier}}]{miles}
{Vazdekis}, A., {S{\'a}nchez-Bl{\'a}zquez}, P., {Falc{\'o}n-Barroso}, J.,
  {et~al.} 2010, \mnras, 404, 1639

\bibitem[{{Venn} {et~al.}(2004){Venn}, {Irwin}, {Shetrone}, {Tout}, {Hill}, \&
  {Tolstoy}}]{Venn04}
{Venn}, K.~A., {Irwin}, M., {Shetrone}, M.~D., {et~al.} 2004, \aj, 128, 1177

\bibitem[{{Viaene} {et~al.}(2018){Viaene}, {Sarzi}, {Zabel}, {Coccato},
  {Corsini}, {Davis}, {De Vis}, {de Zeeuw}, {Falc{\'o}n-Barroso}, {Gadotti},
  {Iodice}, {Lyubenova}, {McDermid}, {Morelli}, {Nedelchev}, {Pinna},
  {Spriggs}, \& {van de Ven}}]{Viaene19}
{Viaene}, S., {Sarzi}, M., {Zabel}, N., {et~al.} 2018, arXiv e-prints
  [\eprint[arXiv]{1812.07582}]

\bibitem[{{Villaume} {et~al.}(2017){Villaume}, {Brodie}, {Conroy},
  {Romanowsky}, \& {van Dokkum}}]{Alexa17}
{Villaume}, A., {Brodie}, J., {Conroy}, C., {Romanowsky}, A.~J., \& {van
  Dokkum}, P. 2017, \apjl, 850, L14

\bibitem[{{Wegner} {et~al.}(2012){Wegner}, {Corsini}, {Thomas}, {Saglia},
  {Bender}, \& {Pu}}]{wegner12}
{Wegner}, G.~A., {Corsini}, E.~M., {Thomas}, J., {et~al.} 2012, \aj, 144, 78

\bibitem[{{Weidner} {et~al.}(2013){Weidner}, {Ferreras}, {Vazdekis}, \& {La
  Barbera}}]{weidner:13}
{Weidner}, C., {Ferreras}, I., {Vazdekis}, A., \& {La Barbera}, F. 2013,
  \mnras, 435, 2274

\bibitem[{{Weilbacher} {et~al.}(2016){Weilbacher}, {Streicher}, \&
  {Palsa}}]{Weilbacher16}
{Weilbacher}, P.~M., {Streicher}, O., \& {Palsa}, R. 2016, {MUSE-DRP: MUSE Data
  Reduction Pipeline}, Astrophysics Source Code Library

\bibitem[{{Wilkinson} {et~al.}(2017){Wilkinson}, {Maraston}, {Goddard},
  {Thomas}, \& {Parikh}}]{Wilkinson17}
{Wilkinson}, D.~M., {Maraston}, C., {Goddard}, D., {Thomas}, D., \& {Parikh},
  T. 2017, \mnras, 472, 4297

\bibitem[{{Worthey}(1994)}]{Worthey94}
{Worthey}, G. 1994, \apjs, 95, 107

\bibitem[{{Worthey} {et~al.}(1992){Worthey}, {Faber}, \&
  {Gonzalez}}]{Worthey92}
{Worthey}, G., {Faber}, S.~M., \& {Gonzalez}, J.~J. 1992, \apj, 398, 69

\bibitem[{{Zhu} {et~al.}(2018{\natexlab{a}}){Zhu}, {van de Ven},
  {M{\'e}ndez-Abreu}, \& {Obreja}}]{Zhu18}
{Zhu}, L., {van de Ven}, G., {M{\'e}ndez-Abreu}, J., \& {Obreja}, A.
  2018{\natexlab{a}}, \mnras, 479, 945

\bibitem[{{Zhu} {et~al.}(2018{\natexlab{b}}){Zhu}, {van den Bosch}, {van de
  Ven}, {Lyubenova}, {Falc{\'o}n-Barroso}, {Meidt}, {Martig}, {Shen}, {Li},
  {Yildirim}, {Walcher}, \& {Sanchez}}]{Ling18}
{Zhu}, L., {van den Bosch}, R., {van de Ven}, G., {et~al.} 2018{\natexlab{b}},
  \mnras, 473, 3000

\bibitem[{{Zieleniewski} {et~al.}(2017){Zieleniewski}, {Houghton}, {Thatte},
  {Davies}, \& {Vaughan}}]{Zieleniewski17}
{Zieleniewski}, S., {Houghton}, R.~C.~W., {Thatte}, N., {Davies}, R.~L., \&
  {Vaughan}, S.~P. 2017, \mnras, 465, 192

\end{thebibliography}

\end{document}